\documentclass[a4paper,11pt]{article}
\usepackage{jheppub} 
\usepackage{lineno}
\usepackage{physics}
\usepackage{subcaption}

\title{\boldmath On the Unruh effect and the thermofield double state}

\author{Gustavo Valdivia-Mera}
\affiliation{Department of Physics, University of Houston,\\
Houston, Texas  77204-5005, USA}

\emailAdd{gvaldiviamera@uh.edu}

\abstract{The purpose of this review is to provide a pedagogical development of the Unruh effect and the thermofield double state. In Section 2, we construct Rindler spacetime and analyze the perspective of an observer undergoing constant acceleration in Minkowski spacetime, which motivates the establishment of the relationship between the Fourier modes in both geometries using the Bogoliubov-Valatin transformation. In Section 3, we explore the underlying physics leading to the Unruh effect, its analogy with the thermal radiation observed around a Schwarzschild black hole, and its manifestation through the coupling of a particle detector to the scalar field. Finally, in Section 4, we derive the thermofield double state by conducting a Euclidean analysis of the field and geometry.}

\begin{document}
\maketitle
\flushbottom

\section{Introduction}
The Unruh effect (also known as the Fulling–Davies–Unruh effect) \cite{fulling1973nonuniqueness,davies1975scalar,Unruh:1976db} describes how an observer in uniform acceleration detects thermal radiation in what an inertial observer would perceive as quantum vacuum. This effect offers significant insights into the behavior of quantum fields in different reference frames, making it evident—via the equivalence principle—its connection with Hawking radiation, and thus showing that the notion of vacuum is observer-dependent. Furthermore, from Israel's work \cite{israel1976thermo}, we understand that the vacuum perception for an observer whose field modes are restricted by causal horizons is thermal, aligning precisely with the thermal description of the ``vacuum" state given by Umezawa and Takahashi \cite{Takahasi:1974zn,Takahashi:1996zn,Umezawa:1982nv,Matsumoto:1982ry}, better known as the thermofield double state. Given all of this, we recognize that the Unruh effect is a highly significant phenomenon, which motivates this pedagogical review.\\

This article is organized as follows:
\begin{itemize}
    \item \textbf{Section 2. Fundamental tools:} In the first part of this section, we address the construction of the Rindler coordinate system by analyzing the dynamics of an observer under constant acceleration. We also examine the extended geometry of Rindler spacetime and conclude by discussing the coordinate transformations for the coordinates in the right and left wedges. In the second part of this section, we analyze the conformal quantum field theory for the massless real scalar field in Minkowski and Rindler geometries, showing how the Fourier modes of the field in both geometries can be related via Bogoliubov transformations.\\
    
    \item \textbf{Section 3. The Unruh effect:} In this section, we analyze the perception of the Minkowski vacuum state from the point of view of an observer undergoing constant acceleration. Additionally, from the equivalence principle, we explore how this analysis connects to gravitational phenomena. Finally, we explore the physics of a particle detector (under constant acceleration) coupled to the scalar field, which provides another manifestation of the Unruh effect.\\
    
    \item \textbf{Section 4. Euclidean approach and the thermofield double state:} In the first part of this section, we demonstrate how the Euclidean path integral approach can be used to obtain the ground state of a quantum field and describe the partition function. Moreover, it is shown that the periodicity of the Euclidean temporal coordinate has a thermal interpretation. In the second part, we develop the construction of the thermofield double state, which characterizes the thermal behavior of the ``vacuum" state and the density matrix of this state.\\

    \item Finally, in \textbf{Section 5}, we present the conclusions, followed by the appendices \textbf{A. Null coordinates}, \textbf{B. Generators of the Lorentz group}, \textbf{C. Properties of the conformal transformation of the metric}, \textbf{D. Pure, mixed, and entangled states}, and \textbf{E. Antiunitary operator}.
\end{itemize}

\section{Fundamental tools}

\subsection{Rindler spacetime}\label{1.1}
\subsubsection{The accelerated observer in Minkowski}\label{1.1.1}
Let $x^{\mu}\equiv(x^0,x^1)$ be the position of an accelerated observer according to an inertial reference frame $\qty{t,x}$\setcounter{footnote}{0}\footnote{Throughout the article we will work in natural units.} in $1+1$ Minkowski spacetime with the line element
\begin{align}\label{dmin}
    ds^2=-dt^2+dx^2.
\end{align}

Let us set $\tau=t\gamma^{-1}$ as the proper time. Then, the spacetime velocity is given by
\begin{align}\label{1}
u^\mu\equiv\left( \frac{dx^0}{d\tau},\frac{dx^1}{d\tau}\right)=\left( \gamma,\gamma v\right),
\end{align}
where $v$ is the spatial velocity, while $\gamma(v)=(1-v^2)^{-\frac{1}{2}}$ is the Lorentz factor. The above expression enables us to obtain the velocity's length\footnote{By length, we are referring to the inner product $x\cdot x=x^\mu x_\mu$.},
\begin{equation}
    u^\mu u_\mu=-1\label{3},
\end{equation}
and the acceleration,
\begin{align}
a^\mu=\frac{du^\mu}{d\tau}\equiv(\gamma^4gv,\gamma^2g+\gamma^4gv^2)\label{2-acc},
\end{align}
where we have considered that the non-inertial system has a constant positive spatial acceleration $(g>0)$. Taking the derivative of \eqref{3} we obtain that the acceleration and velocity are orthogonal:
\begin{equation}
a^\mu u_{\mu}=0\label{av=0}.
\end{equation} 

At each instant of time $t=t_o$ we may consider the non-inertial system as an inertial one with constant velocity $v_o=v(t_o)$ according to $\qty{t,x}$. In addition, in its co-moving frame $(v=0)$ we have $\gamma(v=0)=1$ and $a^\mu(\gamma=1,v=0)=(0 , g)$. Then, from the invariance of the spacetime interval we obtain
\begin{equation}\label{aa=gg}
 a^\mu a_{\mu}=a^0 a_0+a^1 a_1 = g^2.
\end{equation}

Equations \eqref{3}, \eqref{av=0} and \eqref{aa=gg} give us the following system of equations:
\begin{align}
-(u^0)^2+(u^1)^2 &= -1\label{10},\\
-a^0 u^0+a^1 u^1 &= 0\label{11},\\
-(a^0)^2+(a^1)^2 &= g^2\label{12}.
\end{align}

The above system can be solved for $a^\mu$ in terms of $u^\mu$, from which we obtain the following system of coupled equations:
\begin{align}
a^0=\pm g u^1\,\,&\rightarrow\,\, \frac{du^0}{d\tau}\pm g u^1\label{eq2},\\
a^1=\pm g u^0\,\,&\rightarrow\,\, \frac{du^1}{d\tau}\pm g u^0\label{eq1}.
\end{align}

Taking the derivative of each of the above expressions with respect to $\tau$ we find
\begin{align}
    \frac{d^2u^0}{d\tau^2}&=\pm g\frac{du^1}{d\tau}\label{eq4},\\
\frac{d^2u^1}{d\tau^2}&=\pm g\frac{du^0}{d\tau}\label{eq3}.
\end{align}

From \eqref{eq2} in \eqref{eq3} and \eqref{eq1} in \eqref{eq4} we have
\begin{align}
\frac{d^2u^0}{d\tau^2}=g^2u^0,\\
\frac{d^2u^1}{d\tau^2}=g^2u^1.
\end{align}\label{pp22}

Solving the above system, we obtain
\begin{align}
    u^0(\tau)=A_1 e^{g\tau}+A_2 e^{-g\tau}\,\,&\to\,\,\frac{du^0}{d\tau}(\tau)=gA_1 e^{g\tau}-gA_2 e^{-g\tau}\label{u0-de},\\
u^1(\tau)=B_1 e^{g\tau}+B_2 e^{-g\tau}\,\,&\to\,\,\frac{du^1}{d\tau}(\tau)=gB_1 e^{g\tau}-gB_2 e^{-g\tau}\label{u1-de}.
\end{align}

Let us consider that at $\tau=0$ $(t=0)$ the accelerated observer has a spatial velocity equal to zero. Thus,
\begin{align}
v(\tau=0)=0\to\gamma(v(\tau=0))=1.
\end{align}

From \eqref{1} and \eqref{2-acc},
\begin{align}
u^\mu(\tau=0)=u^\mu(\gamma=1,v=0)\,\,&\to\,\, u^\mu(\tau=0)\equiv(1,0)\label{1st-ic},\\
\frac{du^\mu}{d\tau}(\tau=0)=a^\mu(\gamma=1,v=0)\,\,&\to\,\, a^\mu(\tau=0)\equiv(0,g)\label{2nd-ic}.
\end{align}

Applying the initial conditions \eqref{1st-ic} and \eqref{2nd-ic} in \eqref{u0-de} and \eqref{u1-de} we find
\begin{align}
    u^0(\tau=0)=1\rightarrow A_1+A_2=1\,\,&,\,\, a^0(\tau=0)=0\rightarrow gA_1-gA_2=0,\\
u^1(\tau=0)=0\rightarrow B_1+B_2=0\,\,&,\,\, a^1(\tau=0)=0\rightarrow gB_1-gB_2=g.
\end{align}

Then, we have
\begin{align}
A_1=A_2=\frac{1}{2}\quad,\quad B_1=-B_2=\frac{1}{2}.
\end{align}

Therefore,
\begin{align}
u^{0}=\frac{1}{2}(e^{g\tau}+e^{-g\tau})&=\cosh(g\tau),\\
u^{1}=\frac{1}{2}(e^{g\tau}-e^{-g\tau})&=\sinh(g\tau).
\end{align}

Integrating those expressions, we arrive to
\begin{align}
u^0=\frac{dt}{d\tau}\quad&\rightarrow\quad\int_{0}^t dt=\int_0^\tau u^0 d\tau,\\
u^1=\frac{dx}{d\tau}\quad&\rightarrow\quad\int_0^x dx=\int_0^\tau u^1 d\tau.
\end{align}

Thus,
\begin{align}
    t(\tau)&=\frac{1}{g}\sinh{(g \tau)},\label{rindler1}\\
x(\tau)&=\frac{1}{g}\cosh{(g \tau)},\label{rindler2}
\end{align}
where in the expression for $x(\tau)$ we have made the shift $x\rightarrow x+\frac{1}{g}$, such that at $\tau=t=0$ the accelerated observer is located at $x=1/g$.\\

\begin{figure}[t]
 \centering
   \begin{subfigure}{0.4\textwidth}
    \includegraphics[width=\linewidth]{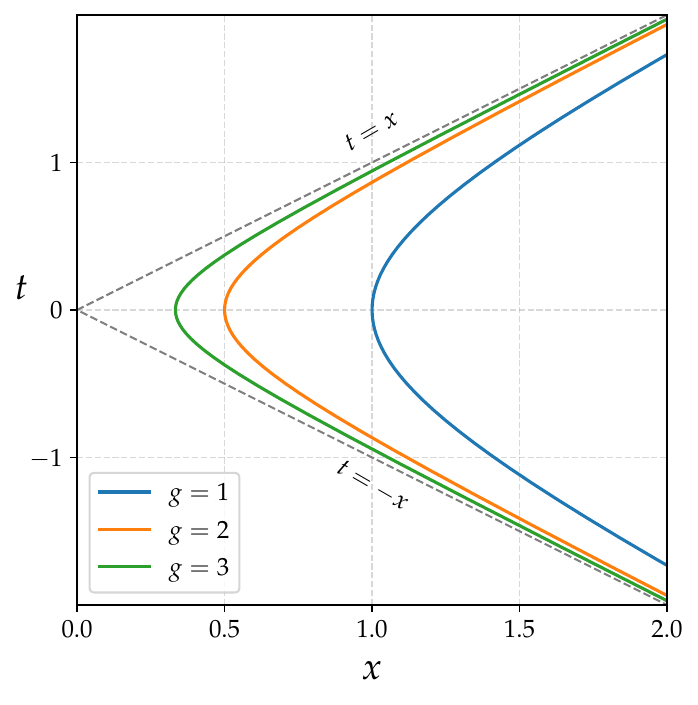}
     \caption{}\label{fig1}
   \end{subfigure}
   \hspace{0.5cm}
   \begin{subfigure}{0.4\textwidth}
     \centering
     \includegraphics[width=\linewidth]{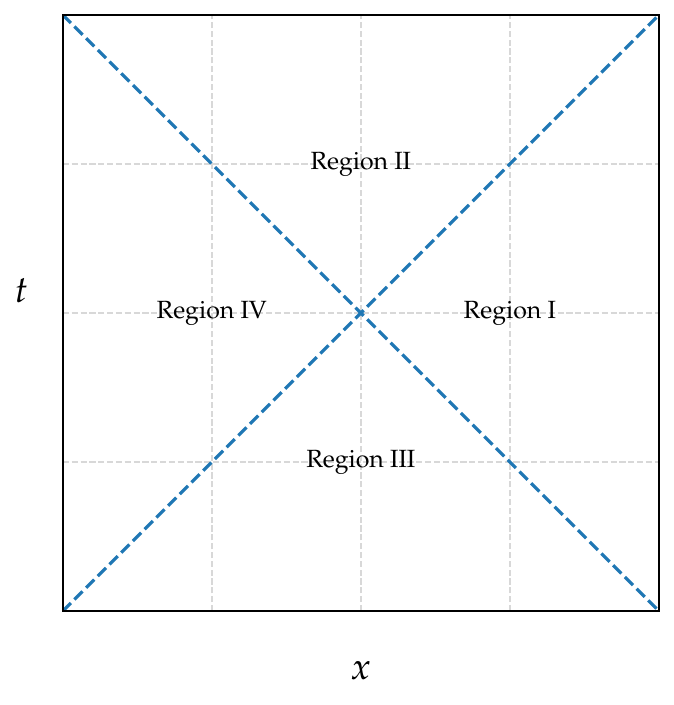}
     \caption{}
\label{fig2}
   \end{subfigure}
   \caption{(a) Trajectories given by \eqref{rindler1} and \eqref{rindler2} (or just \eqref{wuha}) for different values of $g$. (b) Four regions in Minkowski spacetime.}
\end{figure}

The functions $t(\tau)$ and $x(\tau)$ give us the trajectory of the accelerated observer according to the inertial frame. In figure \ref{fig1} we show the parametric curve $(t(\tau),x(\tau))$, which is restricted to the region I of Minkowski spacetime (figure \ref{fig2}). That trajectory can also be written as
\begin{equation}
x=\sqrt{t^2+\frac{1}{g^2}}\label{wuha}.
\end{equation}

In figure \ref{fig2} we show the four regions in Minkowski spacetime, whose boundaries are given by $t=\pm x$. Those regions are defined by the following inequalities:
\begin{align}
\textrm{Region I}&:-x<t<x\label{rI},\\
\textrm{Region II}&:x<t\cap -x<t\label{rII},\\
\textrm{Region III}&:t<x\cap t<-x\label{rIII},\\
\textrm{Region IV}&:x<t<-x\label{rIV}.
\end{align}

Dividing expressions in \eqref{rindler1} and \eqref{rindler2} we obtain that the leaves of foliation for constant $\tau$ are straight lines that pass through the origin. The asymptotic regions $t=\pm x$ for the accelerated observer will be reached at $\tau\rightarrow\pm\infty$ (figure \ref{fig3}), such that
\begin{align}
\frac{t}{x}&=\tanh{(g \tau)}\label{t/x1}.
\end{align}

\subsubsection{Rindler coordinates}\label{1.1.2}
Although the transformations in \eqref{rindler1} and \eqref{rindler2} only give us the trajectory of the accelerated observer, we can use those in order to obtain a complete non-inertial perspective. That is to say, we need a spatial coordinate $\xi$ which, together with $\tau$, enables us to describe any physical phenomena in such a system, which will be called Rindler coordinates.\\

Let us start by expressing the transformations \eqref{rindler1} and \eqref{rindler2} in exponential terms:
\begin{align}
x&=\frac{1}{2}\left(\frac{e^{g\tau}}{g}+\frac{e^{-g\tau}}{g}\label{e+e1}\right),\\
t&=\frac{1}{2}\left(\frac{e^{g\tau}}{g}-\frac{e^{-g\tau}}{g}\right)\label{e-e1}.
\end{align}

Using null coordinates (appendix \ref{Apendice1}) for $\qty{t,x}$, we obtain
\begin{align}
\overline{v}&=t+x\label{cnu1},\\
\overline{u}&=t-x\label{cnu2}.
\end{align}

From \eqref{e+e1} and \eqref{e-e1}, in the previous equations we find
\begin{align}
\overline{v}&=\frac{e^{g\tau}}{g}\label{ue+},\\
\overline{u}&=-\frac{e^{-g\tau}}{g}\label{ue-}.
\end{align}

\begin{figure}[t]
\centering
     \includegraphics[width=0.4\linewidth]{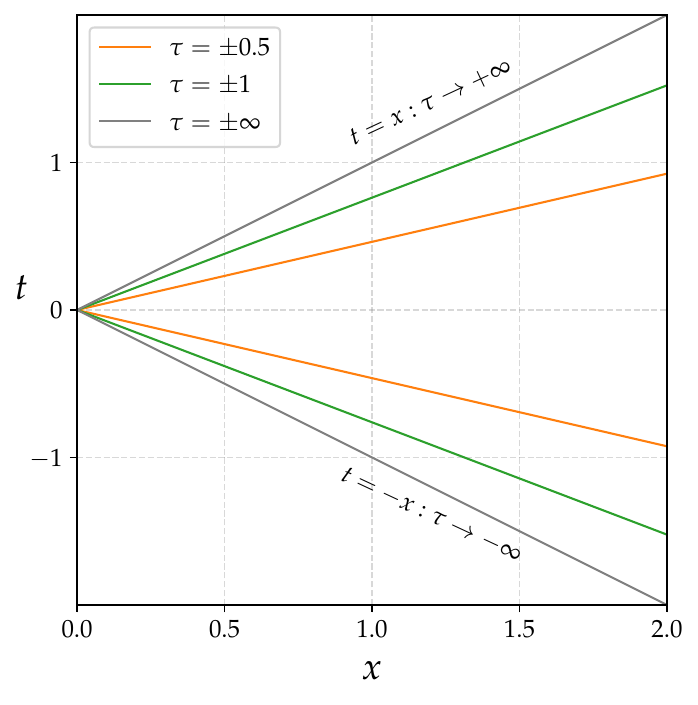}
     \caption{Leaves of foliation for $\tau=\pm 0.5, \pm 1$ and $\pm\infty$ for $g=1$.}\label{fig3}
\end{figure}

Solving $\tau$ from \eqref{ue+} and \eqref{ue-} and adding these terms, we have
\begin{align}\label{tauvu}
\tau=\frac{\frac{1}{g}\ln{(\overline{v}g)}+\left( -\frac{1}{g}\ln{(-\overline{u}g)} \right) }{2}.
\end{align}

Let us set $\xi$ as the spatial coordinate according to the accelerated frame. Therefore, we have the following null coordinates for $\qty{\tau,\xi}$:
\begin{align}
v&=\tau+\xi\label{vxi},\\
u&=\tau-\xi\label{uxi}.
\end{align}

Solving for $\tau$ and $\xi$ we obtain
\begin{align}
\tau &=\frac{v+u}{2}\label{v+u2},\\
\xi &=\frac{v-u}{2}\label{v-u2}.
\end{align}

Associating each null coordinate in Minkowski and Rindler in equations \eqref{tauvu} and \eqref{v+u2}, we find
\begin{align}
\frac{1}{g}\ln{(\overline{v}g)}=v\quad\rightarrow\quad\overline{v}=\frac{1}{g}e^{gv}\label{vv1},\\
-\frac{1}{g}\ln{(-\overline{u}g)}=u\quad\rightarrow\quad\overline{u}=-\frac{1}{g}e^{-gu}\label{uu1}.
\end{align}

Then,
\begin{align}
t+x&=\frac{1}{g}e^{g(\tau+\xi)},\\
t-x&=-\frac{1}{g}e^{-g(\tau-\xi)}.
\end{align}

Solving for $t$:
\begin{align}
t&=\frac{1}{2}\left( \frac{1}{g}e^{g\tau+g\xi}-\frac{1}{g}e^{-g\tau+g\xi} \right)\nonumber\\
&=\frac{e^{g\xi}}{g}\left( \frac{e^{g\tau}-e^{-g\tau}}{2} \right)\label{rd-t}.
\end{align}

Solving for $x$:
\begin{align}
x&=\frac{1}{2}\left( \frac{1}{g}e^{g\tau+g\xi}+\frac{1}{g}e^{-g\tau+g\xi} \right)\nonumber\\
&=\frac{e^{g\xi}}{g}\left( \frac{e^{g\tau}+e^{-g\tau}}{2} \right)\label{rd-x}.
\end{align}

Therefore,
\begin{align}
    t&=\frac{e^{g\xi}}{g}\sinh{(g\tau)},\label{crde000}\\
x&=\frac{e^{g\xi}}{g}\cosh{(g\tau)}.\label{crde0001}
\end{align}

\begin{figure}[t]
\centering
     \includegraphics[width=0.4\linewidth]{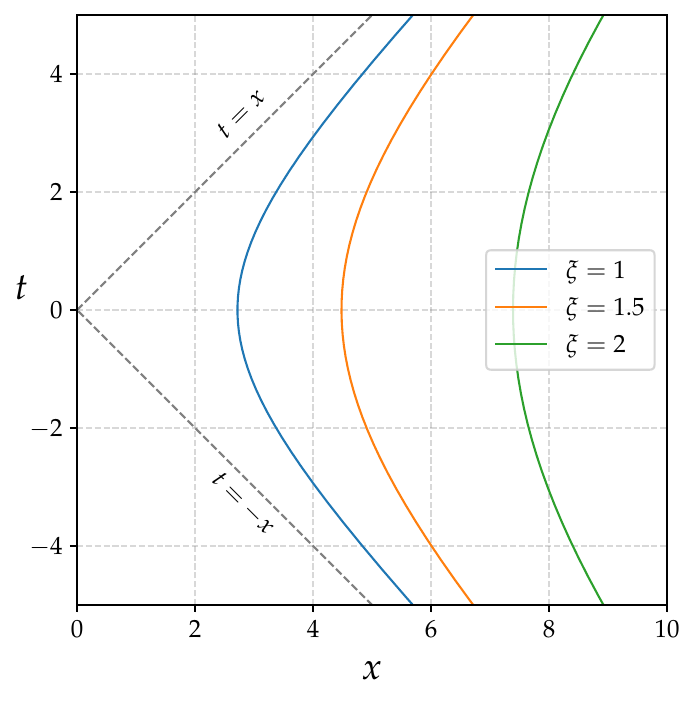}
     \caption{Trajectories given by \eqref{crde000} and \eqref{crde0001} for different values of $\xi$ with $g=1$.}\label{fig4}
\end{figure}

As before, these transformations are defined in the region I of Minkowski spacetime (figure \ref{fig2}). In figure \ref{fig4} we show the parametric curve $(t(\tau,\xi),x(\tau,\xi))$, which can also be written as
\begin{equation}
x=\sqrt{t^2+\frac{e^{2g\xi}}{g^2}}\label{xt2e},
\end{equation}
where the observers at rest (fixed $\xi=\xi_0$) in Rindler spacetime describe a hyperbolic trajectory \eqref{xt2e} in Minkowski. In addition, the leaves of foliation for constant $\tau$ are given by the same expression obtained in \eqref{t/x1}: the asymptotic regions for the accelerated observer ($t=\pm x$) are reached at $\tau\rightarrow\pm\infty$ (figure \ref{fig3}).\\

From \eqref{crde000}, \eqref{crde0001} and \eqref{dmin} we obtain the line element of Rindler spacetime,
\begin{equation}\label{dsrin}
ds^2=e^{2g\xi}(-d\tau^2+d\xi^2).
\end{equation}

The line element \eqref{dsrin} is complete in the sense that we can perform spacetime measurements. In addition, since we have not imposed any restrictions on the Rindler coordinates $(\tau , \xi)$ in the transformations \eqref{crde000} and \eqref{crde0001}, these have the ranges
\begin{align}
-\infty<\tau<+\infty\label{-inftauinf},\\
-\infty<\xi<+\infty\label{-infxiinf}.
\end{align}

Furthermore, the line element \eqref{dsrin} is independent of $\tau$. Thus, $\partial_\tau$ is a Killing vector,
\begin{align}
\partial_\tau&=\frac{\partial t}{\partial \tau}\partial_t+\frac{\partial x}{\partial \tau}\partial_x\nonumber\\
&=g(x\partial_t+t\partial_x)\label{partau},
\end{align}
which generates temporal translations in Rindler spacetime, while acting as a boost in Minkowski (scaled by a factor of $g$). See Appendix \ref{Apendice3} for a review of the generators of the Lorentz group.\\

In addition, it can be expressed as a contravariant vector in Minkowski spacetime,
\begin{equation}\label{bmu}
\tau^\mu\equiv(\tau^0,\tau^1)=(gx, gt),
\end{equation}
with length given by
\begin{equation}\label{bb}
\tau^\mu \tau_\mu=-g^2x^2+g^2t^2=g^2(t+x)(t-x).
\end{equation}

Such that, according to \eqref{rI}, \eqref{rII}, \eqref{rIII} and \eqref{rIV} we obtain
\begin{align}
\text{Region I:}   & \quad \tau^\mu \tau_\mu < 0  \quad \text{(Timelike)},\\
\text{Region II:}  & \quad \tau^\mu \tau_\mu > 0  \quad \text{(Spacelike)},\\
\text{Region III:} & \quad \tau^\mu \tau_\mu > 0  \quad \text{(Spacelike)},\\
\text{Region IV:}  & \quad \tau^\mu \tau_\mu < 0  \quad \text{(Timelike)}.
\end{align}

On $x=\pm\,t$ we have $\tau^\mu \tau_\mu =0$. These regions are called killing horizons and correspond to $\tau\rightarrow\pm\infty$.\\

We can use this killing vector to determine in which direction the temporal coordinate $\tau $ evolves, according to Minkowski. It will confirm not only what we already know from \eqref{crde000} and \eqref{crde0001}, but also an important subtlety in region IV. For that purpose, we use the time basis vector in Minkowski,
\begin{equation}\label{Amu}
\partial_t \,\,\rightarrow\,\, t^\mu\equiv(1,0),
\end{equation}
which has been established, by definition, as future-directed $(t^0>0)$, which is a global behavior due to the flat geometry of Minkowski.\\

As any future-directed vector (timelike or spacelike) has a positive temporal component in its contravariant representation, its product with $t^\mu$ will give us a negative quantity. Therefore, $\tau^\mu$ will be future-directed if $\tau^\mu t_\mu < 0$ and, past-directed if $\tau^\mu t_\mu > 0$. Then, we have
\begin{equation}
\tau^\mu t_\mu = (gx , gt)\cdot(-1,0)=-gx,
\end{equation}
such that in region I, where $x>0$, $\tau$ and $t$ evolve in the same direction, as can be seen from the hyperbolic relation between $t$ and $\tau$ in \eqref{crde000}. On the other hand, in region IV, where $x<0$, $\tau$ evolves in the opposite direction of $t$, the same feature is observed in the extended geometry of Schwarzschild. It suggests that region IV, in the light of the equivalence principle, may be seen as a time-reversed copy of region I. We will approach that behavior in the next section.\\

Finally, let us present an alternative form of the Rindler metric, which is more relevant for studying the near-horizon geometry and its temperature. Let us define
\begin{equation}
    \rho=\frac{e^{g\xi}}{g}\label{rhotogxi}.
\end{equation}

Replacing \eqref{rhotogxi} in \eqref{crde000} and \eqref{crde0001}, we obtain the following map to Minkowski:
\begin{align}
    t&=\rho\sinh{(g\tau)},\label{x-rho-o}\\
x&=\rho\cosh{(g\tau)}.\label{x-rho-o4}
\end{align}

From \eqref{x-rho-o}, \eqref{x-rho-o4} and \eqref{dmin} we obtain an alternative version of the line element \eqref{dsrin} for Rindler spacetime,
\begin{equation}\label{rind-rho-ome}
    ds^2=-\rho^2 g^2d\tau^2+d\rho^2.
\end{equation}

Comparing \eqref{rhotogxi} with \eqref{-inftauinf} and \eqref{-infxiinf} we obtain the following ranges for Rindler coordinates $(\rho , \tau)$:
\begin{align}
0<&\rho<+\infty\label{0rhoinf},\\
-\infty<&\tau<+\infty\label{-infwinf}.
\end{align}

\subsubsection{Extended geometry}\label{caustru}
We have built Rindler spacetime from the analysis developed for the accelerated observer in Minkowski, from which we obtained that Rindler is only a portion of Minkowski, specifically region I. In addition, we obtained that in region IV of Minkowski, $\tau$ evolves in the opposite direction of $t$, thus, one may think that it contains a time-reversed copy of Rindler.\\

In this section, we will assume the existence of Rindler spacetime independently of Minkowski, and we will analyze its extended geometry. Then, in order to reach our goal, we will analyze Rindler in analogy with Kruskal-Szekeres coordinates, and then we will construct the Penrose-Carter diagram.\\

The metric associated with the line element \eqref{rind-rho-ome}, with $\omega = g\tau$,
\begin{equation}\label{rind-rho-ome222}
    ds^2 = -\rho^2 d\omega^2 + d\rho^2,
\end{equation}
is singular at $\rho = 0$ because its determinant vanishes at that point.
\begin{equation}
g_{\mu\nu}\equiv\begin{pmatrix}
-\rho^2 & 0\\
0 & 1
\end{pmatrix}\quad,\quad g=\det g_{\mu\nu}=-\rho^2.
\end{equation}

Then, the inverse metric has a singular point at $\rho= 0$, this is why we imposed $\rho>0$,
\begin{equation}
g^{\mu\nu}\equiv\begin{pmatrix}
-1/\rho^2 & 0\\
0 & 1
\end{pmatrix}.
\end{equation}

To deal with this singularity, which is actually a coordinate singularity, and obtain a better understanding of the causal structure of Rindler spacetime, we need to introduce null coordinates. From the null condition for the line element, $0=-\rho^2d\omega^2+d\rho^2$, we obtain
\begin{equation}\label{tau+-xi}
    \omega=\pm\ln{(\rho)}+const.
\end{equation}

\begin{figure}[t]
\centering
   \begin{subfigure}{0.4\textwidth}
     \includegraphics[width=\linewidth]{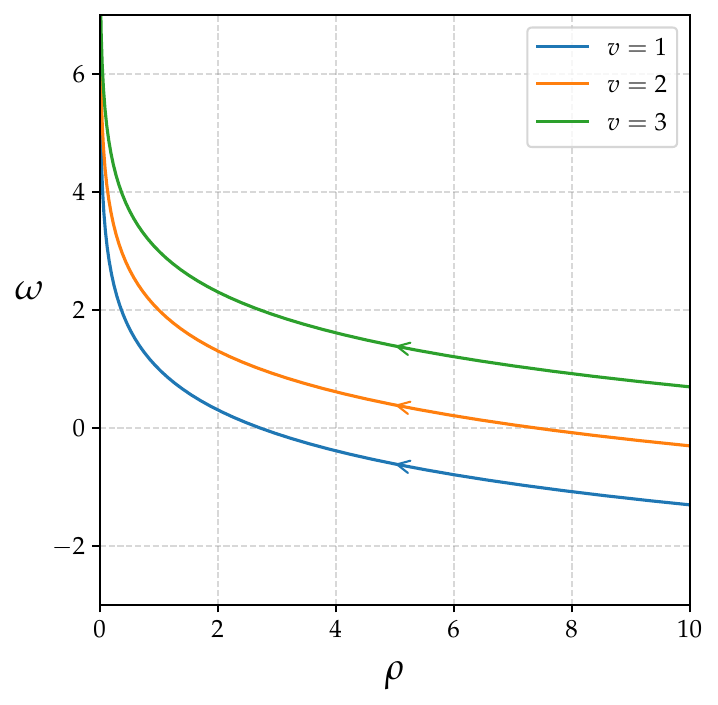}
     \caption{}\label{fig6}
   \end{subfigure}\hspace{0.5cm}
   \begin{subfigure}{0.4\textwidth}
     \includegraphics[width=\linewidth]{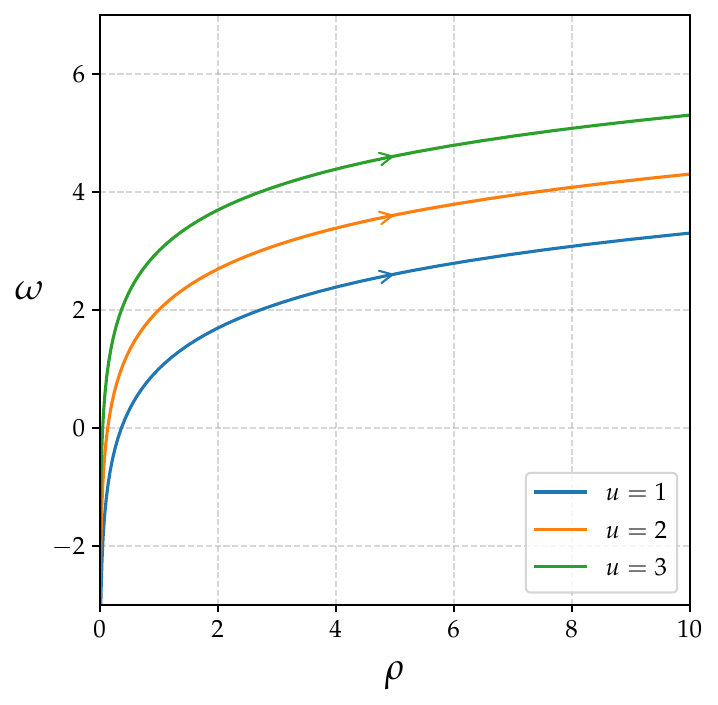}
     \caption{}
\label{fig7}
   \end{subfigure}
   \caption{(a) In-going null geodesics given by different values of $v$ according to \eqref{w+lnrho}. (b) Out-going null geodesics given by different values of $u$ according to \eqref{w-lnrho}.}
\end{figure}

This gives us the null coordinates
\begin{align}
u=\omega-\ln{(\rho)}\label{w-lnrho},\\
v=\omega+\ln{(\rho)}\label{w+lnrho},
\end{align}
where $u$ and $v$ corresponds to out-going and in-going null geodesics respectively, as shown in figures \ref{fig6} and \ref{fig7}.\\

From \eqref{0rhoinf} and \eqref{-infwinf} with \eqref{w-lnrho} and \eqref{w+lnrho} we find
\begin{align}
-\infty<u<+\infty\label{-infuinf},\\
-\infty<v<+\infty\label{-infvinf}.
\end{align}

In terms of the null coordinates $(u,v)$, given in \eqref{w-lnrho} and \eqref{w+lnrho}, the line element \eqref{rind-rho-ome222} takes the form
\begin{equation}
ds^2=-e^{v-u}dudv\label{e-vmetric}.
\end{equation}

In \eqref{e-vmetric}, we still have a singularity in the coordinates due to the exponential term, which falls off when $v-u$ goes to $-\infty$, and precisely corresponds to the singularity at $\rho=0$, with $\omega=\omega_o$ finite, as can be seen from \eqref{w-lnrho} and \eqref{w+lnrho}.\\

Let us return to the line element \eqref{rind-rho-ome222} and check that it is independent of $\omega$, which is the temporal coordinate, then there is a killing vector associated with the symmetry under temporal displacements. This is represented by $\partial_\omega$, which can be expressed as the contravariant vector
\begin{equation}
    \omega^\mu\equiv(1,0).
\end{equation}

In addition, the conserved quantity due to this symmetry is the energy per unit mass\footnote{$\varepsilon$ is known as the energy per unit mass because in Minkowski this expression reduces to $\gamma$, for $E=m\gamma$ (natural units).},
\begin{align}
    \varepsilon&=-g_{\mu\nu}\omega^{\mu}\frac{d\rho^\nu}{d\lambda}\nonumber\\
    &=\rho^2\frac{d\omega}{d\lambda}\label{102},
\end{align}
where $\lambda$ is the affine parameter along the geodesic $\rho^\nu=(\omega,\rho)$. Let us examine the affine parameter for the out-going null geodesic, where $u=\textrm{constant}$. Then, from \eqref{w-lnrho} and \eqref{w+lnrho} we obtain
\begin{align}
    v-u=\ln{(\rho^2)}\quad&\rightarrow\quad \rho^2=e^{v-u}\label{103},\\
    \omega=\frac{u+v}{2}\quad&\rightarrow\quad d\omega=\frac{dv}{2}\label{104}.
\end{align}

From \eqref{103} and \eqref{104} in \eqref{102} we find
\begin{align}
    \int d\lambda = \int \frac{e^{v-u}}{2\varepsilon}dv,\\
    \lambda=\frac{e^{v-u}}{2\varepsilon}\quad\rightarrow\quad \lambda=\left(\frac{e^{-u}}{2\varepsilon}\right)e^v\label{106}.
\end{align}

Since the term inside the parentheses on the right side of \eqref{106} is constant, we set $\lambda_{\textrm{out}}=e^v$ as the affine parameter on the out-going null geodesics.\\

In the same way, for the case of the affine parameter along the in-going geodesics, where $v=\textrm{constant}$, from \eqref{w-lnrho} and \eqref{w+lnrho}, equation \eqref{104} takes the form
\begin{equation}
    \omega=\frac{u+v}{2}\quad\rightarrow\quad d\omega=\frac{du}{2}\label{107}.
\end{equation}

From \eqref{103} and \eqref{107} in \eqref{102} we have
\begin{align}
    \int d\lambda = \int \frac{e^{v-u}}{2\varepsilon}du,\\
    \lambda=-\frac{e^{v-u}}{2\varepsilon}\quad\rightarrow\quad \lambda=-e^{-u}\left(\frac{e^{v}}{2\varepsilon}\right)\label{109}.
\end{align}

Proceeding analogously, we set $\lambda_{\textrm{in}}=-e^{-u}$ as the affine parameter on the in-going null geodesics.\\

From \eqref{-infuinf} and \eqref{-infvinf} we obtain the following ranges for the affine parameters:
\begin{align}
-\infty<\lambda_{\textrm{in}}<0\label{-infubar0},\\
0<\lambda_{\textrm{out}}<+\infty\label{0vbarinf},
\end{align}
which have a barrier in $\lambda_{\textrm{in}}=\lambda_{\textrm{out}}=0$ due to the singularity at $\rho=0$. Then, in order to avoid that singularity, we define a new pair of null coordinates $(\bar{u},\bar{v})$ given by the affine parameters for in-going and out-going null geodesics $(\lambda_{\textrm{in}},\lambda_{\textrm{out}})$,
\begin{align}
    \bar{u}(u)=\lambda_{\textrm{in}}\quad&\rightarrow\quad \bar{u}=-e^{-u}\label{ubar-e-u},\\
    \bar{v}(v)=\lambda_{\textrm{out}}\quad&\rightarrow\quad \bar{v}=e^{v}\label{vbar-e-v}.
\end{align}

From \eqref{-infubar0} and \eqref{0vbarinf}, we know that these null coordinates are restricted to $\bar{u}<0$ and $\bar{v}>0$. However, as the line element \eqref{e-vmetric} is now given by
\begin{equation}\label{ds-barubarv}
ds^2=-d\bar{u}d\bar{v},
\end{equation}
where there is no longer any singularity, we can extend the geometry beyond the constraints imposed at $\bar{u}=\bar{v}=0$ in \eqref{-infubar0} and \eqref{0vbarinf}. Therefore,
\begin{align}
-\infty<&\bar{u}<+\infty\label{ubarrange},\\
-\infty<&\bar{v}<+\infty\label{vbarrange}.
\end{align}

As $(\bar{u},\bar{v})$ are null coordinates, let us define those as
\begin{align}
    \bar{u}&=t-x\label{ubar=t-x},\\
    \bar{v}&=t+x\label{vbar=t+x}.
\end{align}

Then, the line element \eqref{ds-barubarv} is now given by
\begin{align}
    ds^2=-dt^2+dx^2.
\end{align}

This result tells us that Minkowski is the extended geometry of Rindler spacetime, that is to say, Rindler is just a portion of Minkowski. In order to see exactly what this portion is, we need to obtain the transformation between $\qty{t,x}$ and $\qty{\omega,\rho}$. From \eqref{w-lnrho}, \eqref{w+lnrho}, \eqref{ubar=t-x} and \eqref{vbar=t+x} in \eqref{ubar-e-u} and \eqref{vbar-e-v} we obtain
\begin{align}
\bar{u}=-e^{-u}\quad\rightarrow\quad t-x=-e^{-\omega+\ln{(\rho)}},\\
\bar{v}=e^{v}\quad\rightarrow\quad t+x=e^{\omega+\ln{(\rho)}},\\
t=e^{\ln{(\rho)}}\left( \frac{e^{\omega}-e^{-\omega}}{2} \right),\\
x=e^{\ln{(\rho)}}\left( \frac{e^{\omega}+e^{-\omega}}{2} \right).
\end{align}

\begin{figure}[t]
  \centering
   \begin{subfigure}{0.3\textwidth}
     \includegraphics[width=\linewidth]{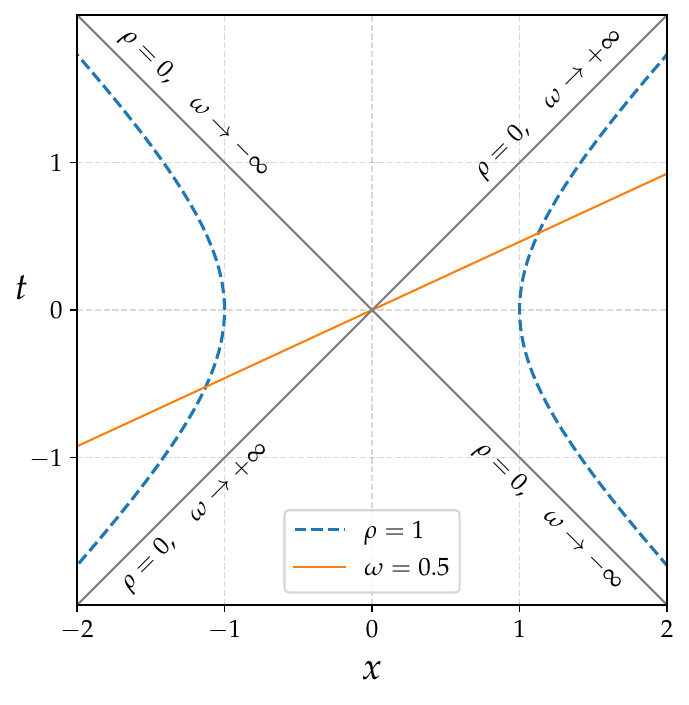}
     \caption{}\label{fig8}
   \end{subfigure}\hspace{.5cm}
   \begin{subfigure}{0.3\textwidth}
     \includegraphics[width=\linewidth]{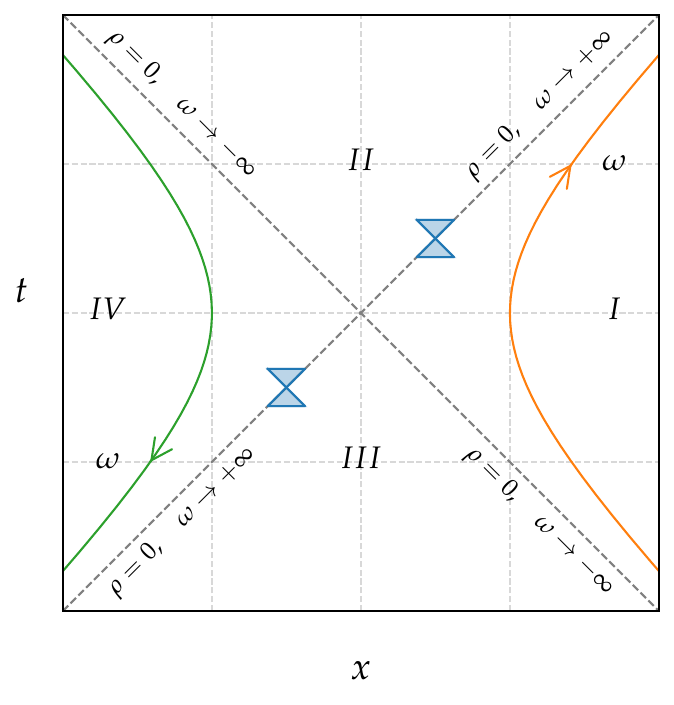}
     \caption{}\label{fig9}
   \end{subfigure}\hspace{.5cm}
   \begin{subfigure}{0.3\textwidth}
     \includegraphics[width=\linewidth]{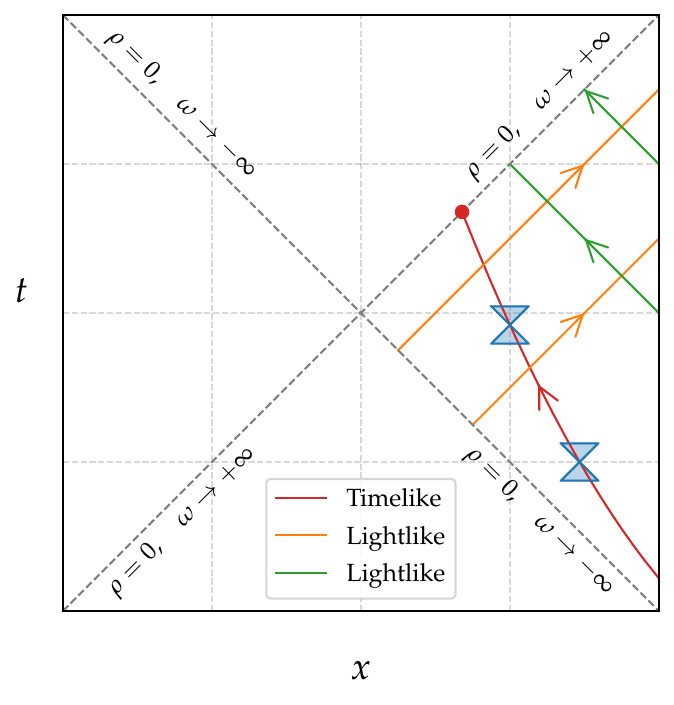}
     \caption{}\label{fig10}
   \end{subfigure}
   \caption{(a) Trajectories for a constant $\rho$ and a leaf of foliation for $\omega$. (b) Causal structure of the extended geometry of Rindler spacetime. (c) Timelike and lightlike geodesics of the Rindler observer in the $tx$-plane.}
\end{figure}

Therefore,
\begin{align}
    t&=\rho\sinh{(\omega)},\label{10-./}\\
x&=\rho\cosh{(\omega)},\label{10-./]}
\end{align}
which is precisely the transformations \eqref{x-rho-o} and \eqref{x-rho-o4}, and, as we already know, it covers the region I of Minkowski spacetime (figure \ref{fig2}). In that sense, the trajectories of a Rindler observer can be seen in the extended geometry as
\begin{align}
    x=\sqrt{t^2+\rho^2}\label{xsqtr},
\end{align}
and the foliations at constant $\omega$ are given by
\begin{align}
\frac{t}{x}=\tanh{(\omega)}\label{toxthan},
\end{align}
such that the asymptotic regions for Rindler observer, given by $t=\pm x$, corresponds to $\omega\rightarrow\pm\infty$ or
$\rho=0\rightarrow\bar{u}=\bar{v}=0$. All these observations are shown in figure \ref{fig8}.\\

Other important features are that the causal structure of the extended geometry of Rindler spacetime is given by $45^{\circ}$ light-cones, which after being introduced give us that regions I and IV are causally disconnected (figure \ref{fig9}). Also, all timelike geodesic remains within the light cones, while lightlike geodesics are represented by $45^{\circ}$ straight lines, as shown in figure \ref{fig10}.\\

As $\omega$ evolves in opposite directions in regions I and IV (figure \ref{fig9}), which is the same feature observed in the extended geometry of Schwarzschild due to Kruskal-Szekeres coordinates, and $x<0$ in region IV, we may say that we have a time-reversed copy of the original Rindler spacetime, as we have deduced in the previous section. Therefore, in region IV the coordinate transformation is given by
\begin{align}
t&=-\rho\sinh{(\omega)},\label{-x-rho-o}\\
x&=-\rho\cosh{(\omega)}.\label{-x-rho-o4}
\end{align}

Moreover, regions II and III are causally analogous to black and white holes, respectively, with horizons on $x=\pm t$, which corresponds to $\omega \rightarrow\pm \infty$ or $\rho=0$ (figure \ref{fig8}).\\

We can summarize the whole extended geometry of Rindler spacetime in a finite-range diagram, namely Penrose-Carter. Using the following maps will allow us to make \textit{finite the infinity},
\begin{align}
\bar{u}&=\tan{\frac{\tilde{u}}{2}}\label{ubar-tan-u/2},\\
\bar{v}&=\tan{\frac{\tilde{v}}{2}}\label{vbar-tan-v/2}.
\end{align}

Which, from the extended geometry expressed in \eqref{ubarrange} and \eqref{vbarrange} gives us the following ranges for $\qty(\tilde{u},\tilde{v})$:
\begin{align}
-\pi< &\tilde{u} < \pi\label{ulala},\\
-\pi< &\tilde{v} < \pi\label{ulele}.
\end{align}

In these new coordinates, the line element \eqref{ds-barubarv} is given by
\begin{equation}
ds^2=-\frac{1}{4}\sec^2{\frac{\tilde{u}}{2}}\sec^2{\frac{\tilde{v}}{2}}d\tilde{u}d\tilde{v}.
\end{equation}

Due to the ranges of $\qty(\tilde{u},\tilde{v})$ in \eqref{ulala} and \eqref{ulele}, we identify the conformal factor
\begin{equation}
    \Omega=\frac{1}{2}\sec{\frac{\tilde{u}}{2}}\sec{\frac{\tilde{v}}{2}}>0.
\end{equation}

Thus,
\begin{equation}
    ds^2=\Omega^2 d\tilde{s}^2.
\end{equation}

\begin{figure}[t]
  \centering
     \includegraphics[width=0.4\linewidth]{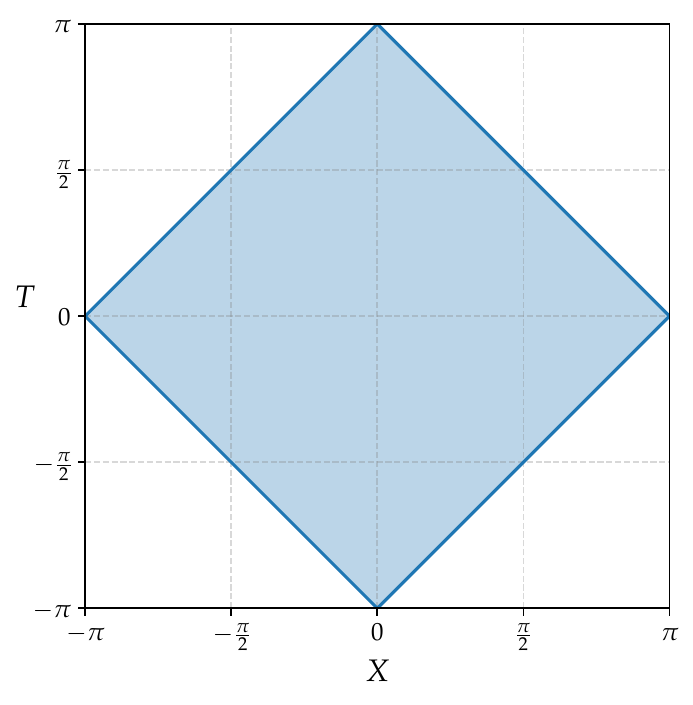}
   \caption{Region delimited by the ranges of $\qty{\tilde{u},\tilde{v}}$ in the $TX$-plane.}\label{TX+Pi-Pi}
\end{figure}

As both metrics are related by the conformal factor $\Omega$, then  null geodesics are preserved (appendix \ref{Apendice2}). Therefore, the metric with line element $ds^2$ has the same causal structure as that with $d\tilde{s}^2$. Defining $\qty(\tilde{u}, \tilde{v})$ as null coordinates,
\begin{align}
\tilde{u}&=T-X,\\
\tilde{v}&=T+X,
\end{align}
we have that the line element \eqref{rind-rho-ome222} can be expressed in a form that is conformal to
\begin{equation}
d\tilde{s}^2= -dT^2+dX^2.
\end{equation}

In consequence, the causal structure is given by $45^{\circ}$ light-cones.\\

\begin{figure}[t]
\centering
     \includegraphics[width=0.8\linewidth]{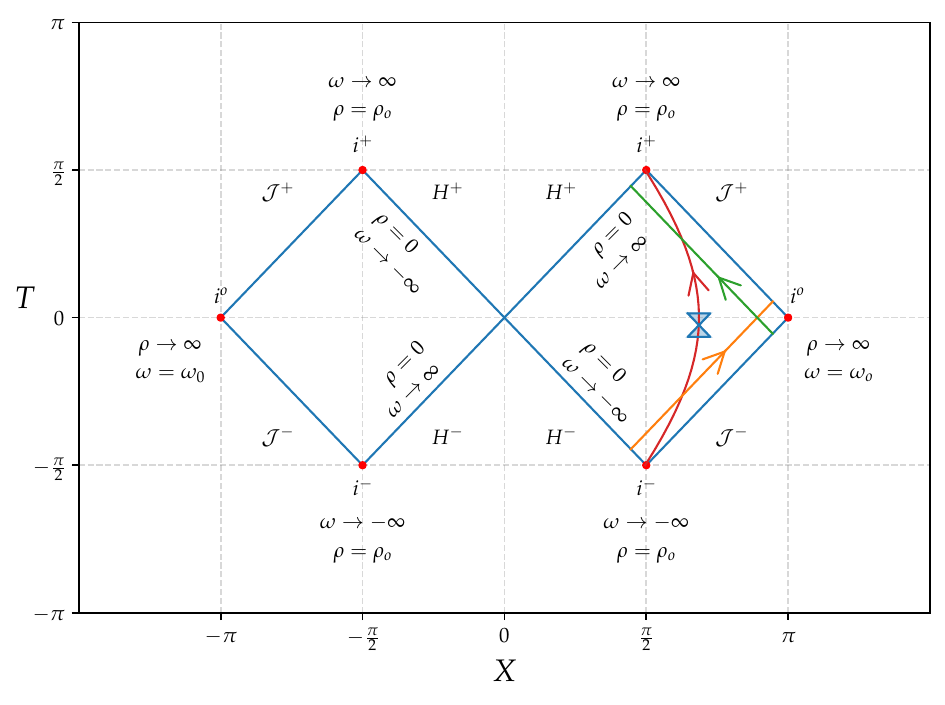}
     \caption{Penrose-Carter diagram of Rindler spacetime.}\label{fig12}
\end{figure}

Now we are able to map the entire extended geometry in a finite region of the $TX$-plane, delimited by \eqref{ulala} and \eqref{ulele} (figure \ref{TX+Pi-Pi}) in what we know as the Penrose-Carter diagram (figure \ref{fig12}). As we know from \eqref{xsqtr}, $\rho=0$ corresponds to $t=\pm x$ , which is the same as $\bar{u}=\bar{v}=0$. Then, from \eqref{ubar-tan-u/2} and \eqref{vbar-tan-v/2}, we obtain that it also corresponds to $\tilde{u}=\tilde{v}=0$, which is the same as $T=\pm X$. Those asymptotic regions can be seen as horizons since the Rindler observer will reach those at $\omega\rightarrow\pm\infty$, as shown in \eqref{toxthan}, and nothing can get out from region II or get into region III. Moreover, the identification of the future $H^+$ and past horizons $H^-$ depends on the null rays, which travel in $45^{\circ}$ straight lines in the $TX$-plane. It means that null rays travel from the past null infinity $\mathcal{J}^{-}$ to the future horizon $H^+$ (green line in figure \ref{fig12}), and from the past horizon $H^-$ to the future null infinity $\mathcal{J}^{+}$ (orange line in figure \ref{fig12}). We also apply this procedure on the copy of Rindler spacetime located on the left side of figure \ref{fig12}.\\

In order to find the future and past timelike infinity, we analyze what happens at $\omega\rightarrow\pm\infty$ (future: $+$, past: $-$) with $\rho=\rho_o$ (finite) in equations \eqref{w-lnrho}, \eqref{w+lnrho}, \eqref{ubar-e-u}, \eqref{vbar-e-v}, \eqref{ubar-tan-u/2} and \eqref{vbar-tan-v/2}:\\

i) Future timelike infinity $i^+$:
\begin{align}
u\rightarrow\infty \quad\rightarrow\quad \bar{u}=0 \quad\rightarrow\quad \tilde{u}=0 \quad\rightarrow\quad T=X,\\
v\rightarrow\infty \quad\rightarrow\quad \bar{v}\rightarrow\infty \quad\rightarrow\quad \tilde{v}=\pi \quad\rightarrow\quad T=-X+\pi.
\end{align}

As both equations must be satisfied, the location of the future timelike infinity is given by the intersection of $\qty{T=X , T=-X+\pi}$: $i^+\equiv(\frac{\pi}{2} , \frac{\pi}{2})$. As we know, on the left side of figure \ref{fig12} we have a copy of Rindler spacetime, then, the future timelike infinity for that copy is located symmetrically: $i^+\equiv(-\frac{\pi}{2} , \frac{\pi}{2})$.\\

ii) Past timelike infinity $i^-$:
\begin{align}
u\rightarrow-\infty \quad\rightarrow\quad \bar{u}=-\infty \quad\rightarrow\quad \tilde{u}=-\pi \quad\rightarrow\quad T=X-\pi,\\
v\rightarrow-\infty \quad\rightarrow\quad \bar{v}=0 \quad\rightarrow\quad \tilde{v}=0 \quad\rightarrow\quad T=-X.
\end{align}

As before, the location of the past timelike infinity is given by the intersection of $\qty{T=X-\pi , T=-X}$: $i^-\equiv(\frac{\pi}{2} , -\frac{\pi}{2})$. In consequence, in the copy of Rindler spacetime, we have $i^-\equiv(-\frac{\pi}{2} , -\frac{\pi}{2})$.\\

Finally, in order to obtain spacelike infinity, $i^o$, we apply $\omega=\omega_o$ (finite) with $\rho\rightarrow\infty$ in equations \eqref{w-lnrho}, \eqref{w+lnrho}, \eqref{ubar-e-u}, \eqref{vbar-e-v}, \eqref{ubar-tan-u/2} and \eqref{vbar-tan-v/2}:
\begin{align}
u\rightarrow-\infty \quad\rightarrow\quad \bar{u}=-\infty \quad\rightarrow\quad \tilde{u}=-\pi \quad\rightarrow\quad T=X-\pi,\\
v\rightarrow\infty \quad\rightarrow\quad \bar{v}\rightarrow\infty \quad\rightarrow\quad \tilde{v}=\pi \quad\rightarrow\quad T=-X+\pi.
\end{align}

Then, the location of the spacelike infinity is given by the intersection of the lines $\qty{T=X-\pi , T=-X+\pi}$: $i^o\equiv(\pi , 0)$. Clearly, in the copy of Rindler spacetime, we have $i^o\equiv(-\pi , 0)$.

\begin{figure}[t]
\centering
     \includegraphics[width=0.4\linewidth]{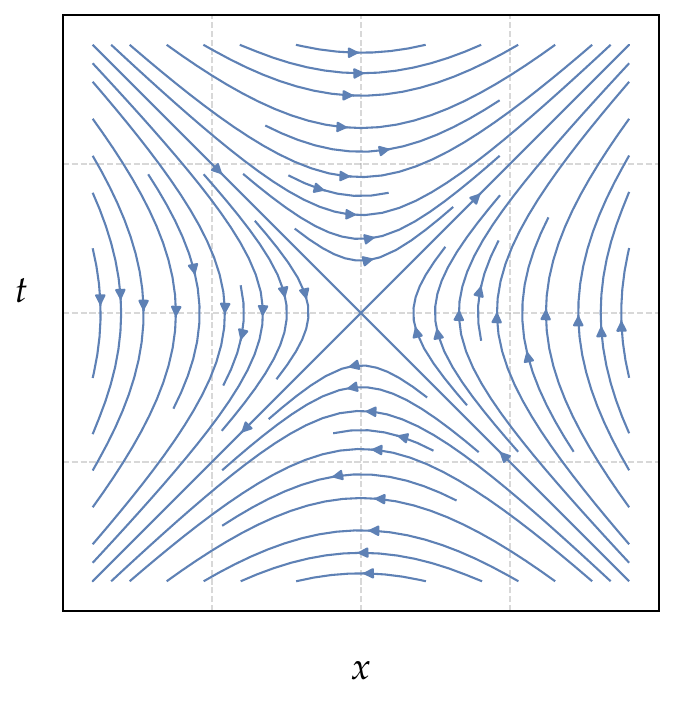}
     \caption{Integral curves of the vector field $\partial_\tau/g$ in Minkowski spacetime.}\label{fig5a}
\end{figure}

\subsubsection{Comments on the coordinate transformations in regions I and IV}

From equations \eqref{crde000},\eqref{crde0001}, \eqref{-x-rho-o} and \eqref{-x-rho-o4}, we have that in regions I and IV the coordinate transformations between Minkowski and Rindler are given by
\begin{align}
t_I = \frac{e^{g\xi}}{g}\sinh{(g\tau)}\quad &, \quad x_I = \frac{e^{g\xi}}{g}\cosh{(g\tau)}, \\
t_{IV} = -t_I\quad &, \quad x_{IV} = -x_I.\label{k10-/.}
\end{align}

Moreover, we note that the line element is in both cases the same and given by \eqref{dsrin},
\begin{equation}
    ds^2 = e^{2g\xi}(-d\tau^2+d\xi^2).
\end{equation}

The Killing vector for this line element, obtained in \eqref{partau},
\begin{equation}
\partial_\tau = g(x\partial_t+t\partial_x),\label{1skill}
\end{equation}
precisely reflects the subtlety of temporal evolution in regions I and IV. This vector, which is the generator of temporal evolution in Rindler spacetime, corresponds to the boost generator in the $x$ direction of Minkowski spacetime, scaled by the factor $g$. In figure \ref{fig5a}, we can observe the integral curves of the vector field $\partial_\tau/g$, where we note that while in region I, temporal evolution is directed towards the future of Minkowski, in region IV it is directed towards the past. It is also appreciated that in regions II and III, the curves are spacelike.\\

\begin{figure}[t]
    \centering
    \begin{subfigure}{0.4\textwidth}
        \includegraphics[width=\linewidth]{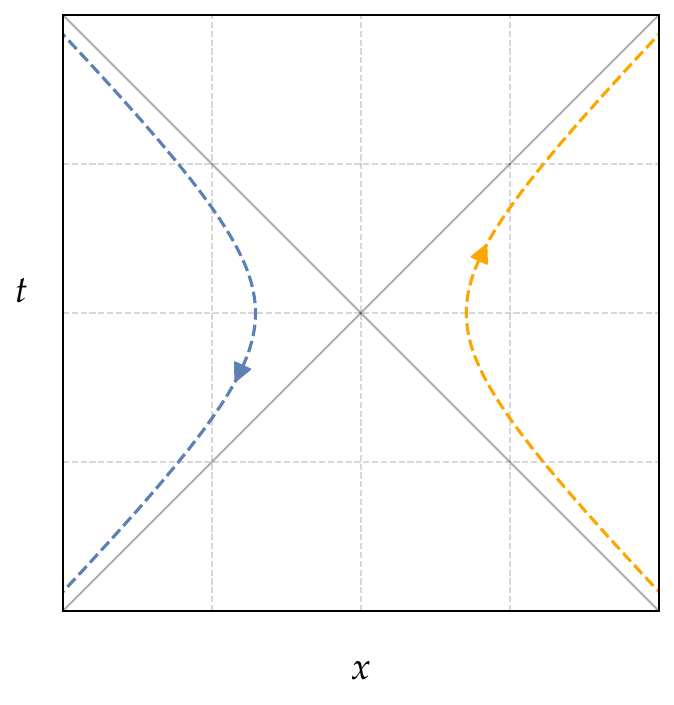}
\captionsetup{justification=centering}
        \caption{}
        \label{fig5b}
    \end{subfigure}
    \hspace{.5cm}
    \begin{subfigure}{0.4\textwidth}
        \includegraphics[width=\linewidth]{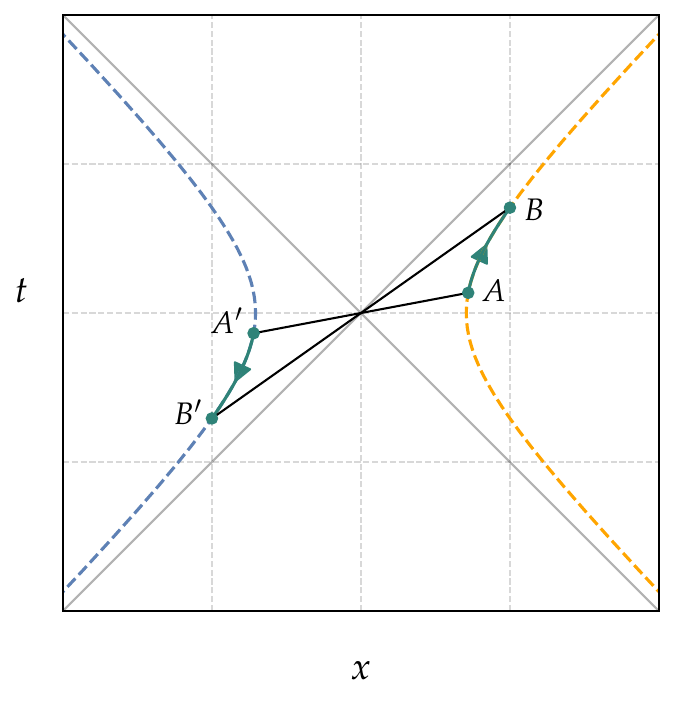}
\captionsetup{justification=centering}
        \caption{}
        \label{fig5c}
    \end{subfigure}    
    \caption{(a) Trajectories for a fixed value of $\xi$. (b) Evolution from $A$ to $B$ related by parity and time-reversal to that from $A'$ to $B'$.}
    \label{fig}
\end{figure}

The trajectories described by the observer under constant acceleration, in regions I and IV, across the entire range of $\tau$ values, describe semi-hyperbolas, which complement each other to form a hyperbola over Minkowski (figure \ref{fig5b}). This hyperbola has null horizons given by $t = \pm x$. Finally, we can observe that temporal evolution on one of the semi-hyperbolas automatically describes the other through parity and time reversal transformations, as shown in figure \ref{fig5c}.

\subsection{Free scalar field and the Bogoliubov-Valatin transformations}
\subsubsection{Scalar field and conformal transformation}\label{sfandct}

\paragraph{Scalar field:} The action for the real massive scalar field in $1+1$ background with metric $g_{\mu\nu}$ is given by
\begin{equation}\label{lagran1}
I=\int d^2x \sqrt{-g}\mathcal{L}=\int d^2x\sqrt{-g}\qty[-\frac{1}{2} g^{\mu\nu}\nabla_{\mu}\phi \nabla_{\nu}\phi-V(\phi)],
\end{equation}
where $V(\phi)=\frac{m^2}{2}\phi^2$. The equation of motion is obtained by taking the variation with respect to the field,
\begin{equation}
    \delta I = \int d^4x\sqrt{-g}\qty[-g^{\mu\nu}\nabla_\mu\phi\delta\nabla_\nu\phi-\delta V\qty(\phi)].
\end{equation}

Using $\qty[\nabla,\delta]=0$ and integrating by parts, we obtain

\begin{equation}
    \delta I = \int d^4x\sqrt{-g}\qty[-\nabla^\mu\qty(\nabla_\mu\phi\delta\phi)+g^{\mu\nu}\qty(\nabla_\mu\nabla_\nu\phi)\delta\phi-\frac{dV}{d\phi}\delta\phi].
\end{equation}

The first term in the bracket is a total derivative, applying Stoke's theorem and the fact that the variation of the field is zero at the boundaries (infinity) we may neglect that term,
\begin{equation}
    \delta I = \int d^4x\sqrt{-g}\qty[g^{\mu\nu}\qty(\nabla_\mu\nabla_\nu\phi)-\frac{dV}{d\phi}]\delta\phi.
\end{equation}

From the variational principle, $\delta I=0$, and $\qty[\nabla_\mu,g^{\mu\nu}]=0$, we find

\begin{equation}
    \nabla_\mu\qty(g^{\mu\nu}\nabla_\nu\phi)=\frac{dV}{d\phi}\label{kg_1}.
\end{equation}

Using the well-known expression of the covariant derivative and the contraction of the Christoffel symbol, we arrive to
\begin{align}
\nabla_\mu\qty(g^{\mu\nu}\nabla_\nu\phi)&=\partial_\mu\qty(g^{\mu\nu}\nabla_\nu\phi)+\Gamma^\mu_{\mu\nu}g^{\nu\alpha}\nabla_\alpha\phi\label{cd_1},\\
    \Gamma^\mu_{\mu\nu}&=\frac{1}{2}g^{\mu\alpha}\partial_\nu g_{\mu\alpha}\label{christo}.
\end{align}

The expression for the Christoffel symbol used above may be written in a different form using the following property for a diagonalizable and invertible matrix $M$,
\begin{equation}\
    \ln\qty(\det M)=\Tr\qty(\ln M).
\end{equation}

Taking the derivative of the above expression
\begin{align}
    \partial_\mu\qty(\ln\qty(\det M))&=\partial_\mu(\Tr\qty(\ln M))\nonumber\\
    \frac{1}{\det M}\partial_\mu\qty(\det M)&=\Tr\partial_\mu(\ln M)\nonumber\\&=\Tr\qty(M^{-1}\partial_\mu M),
\end{align}
and making $M\equiv g_{\mu\nu}$ we have
\begin{equation}
    \frac{1}{g}\partial_\nu g = g^{\mu\alpha}\partial_\nu g_{\mu\alpha}.
\end{equation}

Then, the Christoffel symbol \eqref{christo} is given by
\begin{equation}
    \Gamma^\mu_{\mu\nu}=\frac{1}{2g}\partial_\nu g = \frac{1}{\sqrt{-g}}\partial_\nu\sqrt{-g}\label{cs_1}.
\end{equation}

Finally, from \eqref{kg_1}, \eqref{cd_1} and \eqref{cs_1} we obtain the following form of the equation of motion for the scalar field $\phi$:
\begin{align}
    \partial_\mu\qty(g^{\mu\nu}\nabla_\nu\phi)+\frac{1}{\sqrt{-g}}\partial_\mu\qty(\sqrt{-g})g^{\mu\nu}\nabla_\nu\phi=\frac{dV}{d\phi}\nonumber\\
    \frac{1}{\sqrt{-g}}\partial_\mu\qty(\sqrt{-g}g^{\mu\nu}\nabla_\nu\phi)=\frac{dV}{d\phi}.
\end{align}

\paragraph{Conformal invariance of the action:} A remarkable feature of the action \eqref{lagran1} is that for $m=0$, it is invariant under the conformal transformation of the metric, and the field\footnote{On $1+1$, the conformal weight of the real massless scalar field is zero, $\triangle=0$. Thus, $\tilde{\phi}=\Omega^{\triangle}\phi=\phi$. In addition, the energy-momentum tensor of the real massless scalar field on 1+1 is traceless. Consequently, this theory is scale-invariant (see appendix D of \cite{Wald:1984rg})} (see, for example, \cite{DiFrancesco:1997nk}),
\begin{align}\label{coni}
\tilde{g}_{\mu\nu}(\tilde{x})=\Omega^{2}(x)g_{\mu\nu}(x) \rightarrow \tilde{g}^{\mu\nu}(\tilde{x}) = \Omega^{-2}(x)g^{\mu\nu}(x),\\\sqrt{-\tilde{g}} = \Omega^{2}(x)\sqrt{-g}\quad,\quad \tilde{\phi}(\tilde{x})=\Omega(x)^\Delta\phi(x).
\end{align}

Thus:
\begin{equation}
\tilde{I}=-\frac{1}{2}\int d^2x \sqrt{-\tilde{g}}\tilde{g}^{\mu\nu}\nabla_{\mu}\phi \nabla_{\nu}\phi=-\frac{1}{2}\int d^2x  \Omega^{2}\sqrt{-g}\Omega^{-2}g^{\mu\nu}\nabla_{\mu}\phi \nabla_{\nu}\phi=I.
\end{equation}

From section \ref{1.1}, we know that the Rindler metric is conformal to that in Minkowski, with $\Omega^2=e^{2g\xi}$. Therefore, in Minkowski $\qty{t,x}$ and Rindler $\qty{\tau,\xi}$ we obtain the same form of the equation of motion,
\begin{equation}\label{eom2}
\Box\phi(t,x)=\Box\phi(\tau,\xi)=0.
\end{equation}

\subsubsection{Fourier modes and inner product}
As we know, at the quantum level, we obtain the same equation of motion for the field operator in both Minkowski and Rindler spacetime \eqref{eom2}. Therefore, the real scalar field operator\footnote{For the sake of simplicity, we are not using hats on the operators: $\hat{\phi}=\phi$, $\hat{a}(k)=a(k)$.} $\phi(x^\mu)$, for $x^\mu\equiv(t,x)$ or $x^\mu\equiv(\tau, \xi)$, can be expressed in terms of Fourier modes:
\begin{equation}
\phi(x^\mu)=\int_{-\infty}^{\infty}\frac{dk}{\sqrt{4\pi \omega_k}}(a(k) e^{ik^\mu x_\mu}+a^{\dagger}(k) e^{-ik^\mu x_\mu}),\label{sqf1}
\end{equation}
where $k^\mu\equiv(k^0,k)$, and $k^0=\omega_k=\sqrt{k^2+m^2}=\abs{k}=\omega>0$, since we are considering a massless field, $m=0$ (required for the conformal invariance of the action).\\

As we know, $a^{\dagger}(k)$ and $a(k)$ satisfy the following commutation rules
\begin{align}
\left[ a(k), a^{\dagger}(k')\right] =\delta(k-k')\label{com2},\\
\left[ a(k),a(k')\right]=\left[ a^{\dagger}(k),a^{\dagger}(k')\right]=0\label{com3}.
\end{align}

Splitting the expression \eqref{sqf1} for $k>0$ and $k<0$, we obtain
\begin{multline}
\phi(x^\mu)=\int_{0}^{\infty}\frac{dk}{\sqrt{4\pi \omega_k}}(a(k) e^{ik^\mu x_\mu}+a^{\dagger}(k) e^{-ik^\mu x_\mu})\,+\\+\int_{-\infty}^{0}\frac{dk}{\sqrt{4\pi \omega_k}}(a(k) e^{ik^\mu x_\mu}+a^{\dagger}(k) e^{-ik^\mu x_\mu}).
\end{multline}

Changing $k\rightarrow -k$ in the second integral enables us to express the field operator as
\begin{multline}\label{phifinal}
\phi(x^\mu)=\int_{0}^{\infty}\frac{dk}{\sqrt{4\pi \omega_k}}(a(k) e^{i(-\omega_k t+kx)}+a^{\dagger}(k) e^{-i(-\omega_k t+kx)}\,+\\+b(k) e^{i(-\omega_k t-kx)}+b^{\dagger}(k) e^{-i(-\omega_k t-kx)}),
\end{multline}
where $b(k)=a(-k)$.\\

In addition, due to the conformal invariance, the field operator in Minkowski and Rindler (regions I and IV)\footnote{In both regions we have the same line element conformal to Minkowski. Thus, in each of them, there is a field operator according to \eqref{phi(tauxi)}.} can be expressed as
\begin{equation}\label{phi(tx)}
\phi(t,x)=\int_{0}^{\infty}d\omega(a(\omega)f_{\omega}+a^{\dagger}(\omega) f^{*}_{\omega}+b(\omega)g_{\omega}+b^{\dagger}(\omega) g^{*}_{\omega}),
\end{equation}
\begin{equation}\label{phi(tauxi)}
\phi(\tau,\xi)=\int_{0}^{\infty}d\omega(c(\omega) h_{\omega}+c^{\dagger}(\omega) h^{*}_{\omega}+d(\omega)j_{\omega}+d^{\dagger}(\omega) j^{*}_{\omega}).
\end{equation}

In both fields, the creation and annihilation operators $\qty{a_{\omega}, a^{\dagger}_{\omega}, b_{\omega}, b^{\dagger}_{\omega}, c_{\omega}, c^{\dagger}_{\omega}, d_{\omega}, d^{\dagger}_{\omega}}$ satisfy the commutation rules \eqref{com2} and \eqref{com3}.\\

The Fourier modes in \eqref{phi(tx)} and \eqref{phi(tauxi)} are given by
\begin{align}
f_{\omega}&=\frac{e^{-i\omega(t-x)}}{\sqrt{4\pi \omega}}\label{mfo1},\\
g_{\omega}&=\frac{e^{-i\omega(t+x)}}{\sqrt{4\pi \omega}}\label{mfo2},\\
h_{\omega}&=\frac{e^{-i\omega(\tau-\xi)}}{\sqrt{4\pi \omega}}\label{mfo3},\\
j_{\omega}&=\frac{e^{-i\omega(\tau+\xi)}}{\sqrt{4\pi \omega}}\label{mfo4}.
\end{align}

The modes $f_{\omega}$ and $g_{\omega}$ are identified as right and left modes, respectively (the same for $h_{\omega}$ and $j_{\omega}$). In addition, each of them represents a complete basis for the field operator, and satisfies the Klein-Gordon inner product, as presented by Wald in \cite{Wald:1993kj},
\begin{equation}\label{pikg}
\langle \mathcal{F}_{\omega}, \mathcal{F}_{\omega'} \rangle:=-i\int_\Sigma \sqrt{\sigma}d\Sigma^\mu\qty[\mathcal{F}_{\omega'}\nabla_\mu \mathcal{F}^{*}_{\omega}-\mathcal{F}^{*}_{\omega}\nabla_\mu \mathcal{F}_{\omega'}],
\end{equation}
where $d\Sigma^\mu=d\Sigma n^\mu$, and $\Sigma$ is a Cauchy surface.\\

Considering $\Sigma$ given by $S=x^0-x_\circ^0=0$, the normal vector $n^\mu$ (see for example    appendix D of \cite{Carroll:2004st})  is defined as
\begin{align}
n^\mu &= -\frac{\zeta^\mu}{\abs{\zeta^\mu\zeta_\mu}^{1/2}},\\
\zeta^\mu  &= g^{\mu\nu}\nabla_\nu S.
\end{align}

For Minkowski \eqref{dmin} and Rindler metric \eqref{dsrin} we obtain
\begin{align}
\textrm{Minkowski}: &\quad n^\mu =\delta^\mu_0\,\, , \,\, \sqrt{\sigma}=1\,\, , \,\, d\Sigma=dx\rightarrow \sqrt{\sigma}d\Sigma^\mu=dx\delta^\mu_0\label{dfsv1},\\
\textrm{Rindler}: &\quad n^\mu =e^{-g\xi}\delta^\mu_0\,\, , \,\, \sqrt{\sigma}=e^{g\xi}\,\, , \,\, d\Sigma=d\xi\rightarrow \sqrt{\sigma}d\Sigma^\mu=d\xi\delta^\mu_0\label{dfsv2}.
\end{align}

From the inner product \eqref{pikg}, the differential surface vectors \eqref{dfsv1} and \eqref{dfsv2}, and the Fourier modes \eqref{mfo1}, \eqref{mfo2}, \eqref{mfo3} and \eqref{mfo4}, we find
\begin{align}
\langle f_{\omega}, f_{\omega'} \rangle=\langle g_{\omega}, g_{\omega'} \rangle&=\delta(\omega-\omega'),\\
\langle f^{*}_{\omega}, f_{\omega'} \rangle=
\langle g^{*}_{\omega}, g_{\omega'} \rangle&=0,\\
\langle h_{\omega}, h_{\omega'} \rangle=\langle j_{\omega}, j_{\omega'} \rangle&=\delta(\omega-\omega'),\\
\langle h^{*}_{\omega}, h_{\omega'} \rangle=
\langle j^{*}_{\omega}, j_{\omega'} \rangle&=0.
\end{align}

The left and right modes are orthogonal to each other. Thus,
\begin{align}
\langle f_{\omega}, g_{\omega'} \rangle=\langle f^{*}_{\omega}, g_{\omega'} \rangle=0\label{desa},\\
\langle h_{\omega}, j_{\omega'} \rangle=\langle h^{*}_{\omega}, j_{\omega'} \rangle=0\label{desa_extra}.
\end{align}

Furthermore, the complex conjugate of the inner product \eqref{pikg} satisfies
\begin{equation}\label{pro1}
{\langle \mathcal{F}_{\omega}, \mathcal{G}_{\omega'} \rangle}^{*}=-\langle \mathcal{F}^{*}_{\omega}, \mathcal{G}^{*}_{\omega'} \rangle=\langle \mathcal{G}_{\omega'}, \mathcal{F}_{\omega} \rangle.
\end{equation}

\subsubsection{Bogoliubov-Valatin transformation}
Let us define the null coordinates $\qty{\bar{u},\bar{v},u,v}$ in Minkowski $\qty{t,x}$ and Rindler $\qty{\tau,\xi}$,
\begin{align}
\bar{u}&=t-x,\\
\bar{v}&=t+x,\\
u&=\tau-\xi,\\
v&=\tau+\xi.
\end{align}

Then, the equations of motion \eqref{eom2} are now expressed as
\begin{align}
\partial_{\bar{u}}\partial_{\bar{v}}\phi &= 0\label{eom5},\\
\partial_{u}\partial_{v}\phi &= 0\label{eom6},
\end{align}
for the scalar field in Minkowski and Rindler null coordinates, respectively.\\

Evidently, by virtue of \eqref{eom5} and \eqref{eom6}, each solution is of the form
\begin{align}
\phi(\bar{u},\bar{v})=\phi(\bar{u})+\phi(\bar{v})\label{phi(uv)},\\
\phi(u,v)=\phi(u)+\phi(v)\label{phi(uvr)}.
\end{align}

It means that $\phi(\bar{u})$ and $\phi(\bar{v})$ are decoupled, which is in agreement with what we obtained in \eqref{desa} and \eqref{desa_extra}. Let us write the decoupled Fourier modes, \eqref{mfo1} and \eqref{mfo2}, in null coordinates,
\begin{align}
f_{\omega}=\frac{e^{-i\omega \bar{u}}}{\sqrt{4\pi \omega}}\label{fmof},\\
g_{\omega}=\frac{e^{-i\omega \bar{v}}}{\sqrt{4\pi \omega}}.
\end{align}

Expanding $\phi(\bar{u})$ and $\phi(\bar{v})$ in those bases we obtain
\begin{align}
\phi(\bar{u})=\int_{0}^{\infty}d\omega(a(\omega) f_{\omega}+a^{\dagger}(\omega) f^{*}_{\omega})\label{phi(ubar)},\\
\phi(\bar{v})=\int_{0}^{\infty}d\omega(b(\omega)g_{\omega}+b^{\dagger}(\omega) g^{*}_{\omega}).
\end{align}

Adding the above expressions, we obtain \eqref{phi(uv)}, which, as expected, is equal to \eqref{phi(tx)}.\\

The same procedure is applied to Rindler modes \eqref{mfo3} and \eqref{mfo4},
\begin{align}
h_{\omega}=\frac{e^{-i\omega u}}{\sqrt{4\pi \omega}}\label{fmoh},\\
j_{\omega}=\frac{e^{-i\omega v}}{\sqrt{4\pi \omega}}.
\end{align}

From \eqref{phi(uvr)} we obtain the decoupled fields
\begin{align}
\phi(u)=\int_{0}^{\infty}d\omega(c(\omega) h_{\omega}+c^{\dagger}(\omega) h^{*}_{\omega}),\\
\phi(v)=\int_{0}^{\infty}d\omega(d(\omega)j_{\omega}+d^{\dagger}(\omega) j^{*}_{\omega}).
\end{align}

As before, the sum of them reproduce \eqref{phi(uvr)} and \eqref{phi(tauxi)}.\\

The right modes in Minkowski and Rindler are related by the Bogoliubov-Valatin transformation \cite{doi:10.1002/prop.19580061102,Valatin:1958ja},
\begin{equation}\label{bogo1}
h_{\omega '}=\int_{0}^{\infty}d\Omega (\alpha_{\omega '\Omega}f_{\Omega}+\beta_{\omega '\Omega}f^{*}_{\Omega}).
\end{equation}

The coefficients $\alpha_{\omega'\Omega}$ and $\beta_{\omega'\Omega}$ are obtained by the Klein-Gordon inner product,
\begin{align}
\langle h_{\omega'}, f_{\omega} \rangle&=\alpha_{\omega'\omega},\\
\langle h_{\omega'}, f^{*}_{\omega} \rangle&=-\beta_{\omega'\omega}.
\end{align}

Applying the property \eqref{pro1} we find
\begin{align}
\langle f_{\omega}, h_{\omega'} \rangle&=\alpha^{*}_{\omega'\omega},\\
\langle f_{\omega}, h^{*}_{\omega'} \rangle&=\beta_{\omega'\omega}.
\end{align}

Thus, we can write the right mode in Minkowski in terms of that in Rindler,
\begin{equation}\label{bogo2}
f_{\omega}=\int_{0}^{\infty}d\Omega (\alpha^{*}_{\Omega\omega}h_{\Omega}-\beta_{\Omega\omega}h^{*}_{\Omega}).
\end{equation}

From the equality between the right modes in Minkowski and Rindler $\qty{\phi(\bar{u}), \phi(u)}$, and since $\bar{u} = \bar{u}(u)$ (see equation \eqref{uu1}), we can express $\phi(\bar{u})$ in terms of $u$, which is precisely $\phi(u)$ (the same analysis applies for $\qty{\phi(\bar{v}), \phi(v)}$), we have
\begin{equation}\label{min=rin}
\int_{0}^{\infty} d\omega (a(\omega)f_{\omega}+a^{\dagger}(\omega)f^{*}_{\omega})=\int_{0}^{\infty} d\omega'(c(\omega')h_{\omega'}+c^{\dagger}(\omega')h^{*}_{\omega'}).
\end{equation}

Then, using \eqref{bogo2} in \eqref{min=rin}, we find
\begin{multline*}
\int_{0}^{\infty} d\omega (a(\omega)f_{\omega}+a^{\dagger}(\omega)f^{*}_{\omega})=\int_{0}^{\infty} d\omega a(\omega)\displaystyle\int_{0}^{\infty}d\Omega (\alpha^{*}_{\Omega\omega}h_{\Omega}-\beta_{\Omega\omega}h^{*}_{\Omega})\,+\\+\int_{0}^{\infty} d\omega a^{\dagger}(\omega)\displaystyle\int_{0}^{\infty}d\Omega (\alpha_{\Omega\omega}h^*_{\Omega}-\beta^*_{\Omega\omega}h_{\Omega}).
\end{multline*}

Rearranging and comparing term by term with the right-hand side of \eqref{min=rin},
\begin{multline*}
\int_{0}^{\infty} d\Omega \displaystyle\int_{0}^{\infty} d\omega (\alpha^{*}_{\Omega\omega} a(\omega)- \beta^*_{\Omega\omega}a^{\dagger}(\omega))h_{\Omega}\,+\\+\int_{0}^{\infty} d\Omega\displaystyle\int_{0}^{\infty} d\omega(\alpha_{\Omega\omega}a^{\dagger}(\omega)-\beta_{\Omega\omega}a(\omega))h^{*}_{\Omega}=\int_{0}^{\infty} d\Omega(c(\Omega)h_{\Omega}+c^{\dagger}(\Omega)h^{*}_{\Omega}),
\end{multline*}
we obtain the annihilation operator in Rindler in terms of the annihilation and creation operators in Minkowski,
\begin{align}\label{ctoa}
c(\Omega)=\int_{0}^{\infty}d\omega (\alpha^{*}_{\Omega\omega}a(\omega)-\beta^{*}_{\Omega\omega}a^{\dagger}(\omega)).
\end{align}

Similarly, using \eqref{bogo1} in \eqref{min=rin}, we obtain the analogous result
\begin{align}
a(\omega)=\int_{0}^{\infty}d\Omega (\alpha_{\Omega\omega}c(\Omega)+\beta^{*}_{\Omega\omega}c^{\dagger}(\Omega)).
\end{align}

Finally, using \eqref{bogo2} in \eqref{bogo1},
\begin{multline}
h_{\omega '}=\int_{0}^{\infty}d\Omega \alpha_{\omega '\Omega}\displaystyle\int_{0}^{\infty}d\omega (\alpha^{*}_{\omega\Omega}h_{\omega}-\beta_{\omega\Omega}h^{*}_{\omega})\,+\\
+\int_{0}^{\infty}d\Omega \beta_{\omega '\Omega}\displaystyle\int_{0}^{\infty}d\omega (\alpha_{\omega\Omega}h^*_{\omega}-\beta^*_{\omega\Omega}h_{\omega}),
\end{multline}
we obtain the following properties for the Bogoliubov-Valatin coefficients:
\begin{align}
\int_{0}^{\infty}d\Omega(\alpha_{\omega'\Omega}\alpha^{*}_{\omega\Omega}-\beta_{\omega'\Omega}\beta^{*}_{\omega\Omega})&=\delta(\omega-\omega')\label{bogopro},\\
\alpha_{\omega'\Omega}\beta_{\omega\Omega}&=\beta_{\omega'\Omega}\alpha_{\omega\Omega}.
\end{align}

\section{The Unruh effect}\label{EfectoUnsubse}
\subsection{``Thermal vacuum" for the accelerated observer}\label{tervcuuaob}
As we saw in the previous section, we can express the creation and annihilation operators of the field in Rindler spacetime in terms of those in Minkowski \eqref{ctoa}. Therefore, the number operator for an arbitrary energy $\omega$ for right modes in the right wedge\footnote{The same procedure is applied to the left wedge (region IV) and left modes.} (region I) of the Rindler spacetime,
\begin{equation}
N_R(\omega)=c^{\dagger}(\omega)c(\omega),
\end{equation}
acts on the energy states of the field in Minkowski. In what follows, we will explore the perception of the vacuum state of the field in Minkowski spacetime from the perspective of an observer under constant acceleration,
\begin{equation}\label{omNrom}
\mel{0_M}{N_R(\omega)}{0_M}=\mel{0_M}{c^{\dagger}(\omega)c(\omega)}{0_M}.
\end{equation}

Using \eqref{ctoa} in the above expression, we obtain
\begin{multline}\label{157}
\mel{0_M}{N_R(\omega)}{0_M}=\bra{0_M}\int_{0}^{\infty}d\Omega' (\alpha_{\omega\Omega'}a^{\dagger}(\Omega')-\beta_{\omega\Omega'}a(\Omega'))\,\times\\ \times\int_{0}^{\infty}d\Omega (\alpha^{*}_{\omega\Omega}a(\Omega)-\beta^{*}_{\omega\Omega}a^{\dagger}(\Omega))\ket{0_M}
\end{multline}
Applying the action of the creation and annihilation operators in Minkowski over the vacuum state $\ket{0_M}$,
\begin{align}
a^{\dagger}(\Omega)\ket{0_M}&=\ket{(1,\Omega)_M},\\
a(\Omega)\ket{0_M}&=0,
\end{align}
we obtain the following expression:
\begin{equation}\label{NN45}
\mel{0_M}{N_R(\omega)}{0_M}=\qty[\int_{0}^{\infty}d\Omega'\bra{(1,\Omega')_M}\beta_{\omega\Omega'}]\qty[\int_{0}^{\infty}d\Omega\beta^*_{\omega\Omega}\ket{(1,\Omega)_M}].
\end{equation}

Thus, what we obtain in \eqref{NN45} is just an infinite sum of $\abs{\beta_{\omega\Omega}}^2$, which is a positive quantity, over the whole energy spectrum $\Omega$ of the field in Minkowski:
\begin{equation}\label{NN}
\mel{0_M}{N_R(\omega)}{0_M}=\int_{0}^{\infty}d\Omega \abs{\beta_{\omega\Omega}}^{2}.
\end{equation}

Then, it is expected to obtain a divergent value.\\

On the other hand, if we made the computation in \eqref{omNrom} using the number operator with energy $\omega$ (for the right modes) in Minkowski, $N_M(\omega)$, instead of $N_R(\omega)$, we find
\begin{equation}\label{asdkjflkajhf}
\mel{0_M}{N_M(\omega)}{0_M}=\mel{0_M}{a^{\dagger}(\omega)a(\omega)}{0_M}=0.
\end{equation}

As we see, the difference between \eqref{NN} and \eqref{asdkjflkajhf} is due to the factor $\beta_{\Omega\omega}$, which is responsible for combining modes of positive and negative norm $\qty{f_\Omega,f^*_\Omega}$ \eqref{bogo1}. Thus, when $\beta_{\Omega\omega}=0$ the Rindler modes with positive and negative norm $\qty{h_{\omega'},h^*_{\omega '}}$ are just a superposition of their analogues in Minkowski,
\begin{align}
h_{\omega '}&=\int_{0}^{\infty}d\Omega (\alpha_{\omega '\Omega}f_{\Omega}),\\
h^*_{\omega '}&=\int_{0}^{\infty}d\Omega (\alpha^*_{\omega '\Omega}f^*_{\Omega}).
\end{align}

So, it is important to know $\beta_{\Omega\omega}$ to understand its physical implication.\\

Let us start from \eqref{min=rin}, and using the values of $f_{\omega}$ and $h_{\omega'}$ given in \eqref{fmof} and \eqref{fmoh}, we obtain
\begin{multline}\label{mirinfo1}
\int_{0}^{\infty} d\omega \qty[a(\omega)\frac{e^{-i\omega\bar{u}}}{\sqrt{4\pi\omega}}+a^{\dagger}(\omega)\frac{e^{i\omega\bar{u}}}{\sqrt{4\pi\omega}}]= \int_{0}^{\infty} d\omega'\qty[c(\omega')\frac{e^{-i\omega' u}}{\sqrt{4\pi\omega'}}+c^{\dagger}(\omega')\frac{e^{i\omega'u}}{\sqrt{4\pi\omega'}}].
\end{multline}
From the coordinate transformation \eqref{uu1} between Minkowski and Rindler we know that both sides of the expression \eqref{mirinfo1} are functions of $u$. In that sense, we can apply the Fourier transformation in order to get them in terms of the energy $\Omega$,
\begin{multline}\label{mirinfo2}
\int_{-\infty}^{\infty}\frac{du}{\sqrt{2\pi}}e^{i\Omega u}\int_{0}^{\infty} d\omega \qty[a(\omega)\frac{e^{-i\omega\bar{u}}}{\sqrt{4\pi\omega}}+a^{\dagger}(\omega)\frac{e^{i\omega\bar{u}}}{\sqrt{4\pi\omega}}]=\\ =\int_{-\infty}^{\infty}\frac{du}{\sqrt{2\pi}}e^{i\Omega u}\int_{0}^{\infty} d\omega'\qty[c(\omega')\frac{e^{-i\omega' u}}{\sqrt{4\pi\omega'}}+c^{\dagger}(\omega')\frac{e^{i\omega'u}}{\sqrt{4\pi\omega'}}].
\end{multline}
Rearranging the right-hand side of the above expression,
\begin{equation}
\int_{0}^{\infty} \frac{d\omega'}{\sqrt{2\omega'}}\qty[ c(\omega')\int_{-\infty}^{\infty}\frac{du}{2\pi} e^{iu(\Omega-\omega')}+c^{\dagger}(\omega')\int_{-\infty}^{\infty}\frac{du}{2\pi} e^{iu(\Omega+\omega')}],
\end{equation}
and using the definition of the Dirac delta function,
\begin{equation}
\delta(\Omega-\omega')=\int_{-\infty}^{\infty} \frac{du}{2\pi}e^{iu(\Omega-\omega')},
\end{equation}
from \eqref{mirinfo2} we find
\begin{multline}
c(\Omega)=\int_{0}^{\infty}d\omega\left[ \qty(\int_{-\infty}^{\infty}\frac{du}{2\pi}\sqrt{\frac{\Omega}{\omega}}e^{i(\Omega u-\omega\bar{u})})a(\omega)+\qty(\int_{-\infty}^{\infty}\frac{du}{2\pi}\sqrt{\frac{\Omega}{\omega}}e^{i(\Omega u+\omega\bar{u})})a^{\dagger}(\omega)\right].
\end{multline}

Comparing term by term with \eqref{ctoa}, and replacing the value of $\overline{u}(u)$ given in \eqref{uu1}, we obtain the following expressions for the Bogoliubov-Valatin coefficients:
\begin{align}
\alpha^{*}_{\Omega\omega}&=\sqrt{\frac{\Omega}{\omega}}\int_{-\infty}^{\infty}\frac{du}{2\pi}e^{i(\Omega u+\frac{\omega}{g}e^{-gu})}\label{aint},\\
\beta^{*}_{\Omega\omega}&=-\sqrt{\frac{\Omega}{\omega}}\int_{-\infty}^{\infty}\frac{du}{2\pi}e^{i(\Omega u-\frac{\omega}{g}e^{-gu})}\label{bint}.
\end{align}

The integrals on the right-hand side of the \eqref{aint} and \eqref{bint} can be solved by doing
\begin{equation}
    \frac{\omega}{g}e^{-gu}= \gamma ie^x,
\end{equation}
where $\gamma=\pm 1$. Thus, we have
\begin{equation}
    u=-\frac{1}{g}\ln{\qty(\frac{g}{\omega})}-\frac{\gamma i\pi}{2g}-\frac{x}{g},
\end{equation}
where we have used that $\gamma i = e^{\gamma i \frac{\pi}{2}}$. Then, we have $du = -dx/g$, and $u\rightarrow\pm \infty$ for $x\rightarrow -\gamma i\pi/2\mp\infty$. Furthermore, since $\gamma = \pm 1$, we have $\gamma \frac{\omega}{g} e^{-gu} = \gamma^2 i e^x = i e^x$. Thus, the integrals \eqref{aint} and \eqref{bint} can be expressed as
\begin{equation}
    \int_{-\infty}^{\infty}\frac{du}{2\pi}e^{i(\Omega u+\gamma\frac{\omega}{g}e^{-gu})}=\frac{e^{\frac{\gamma \pi \Omega}{2g}}}{2\pi g}\qty(\frac{g}{\omega})^{-\frac{i\Omega}{g}}\int_{-\frac{\gamma i\pi}{2}-\infty}^{-\frac{\gamma i\pi}{2}+\infty}dx e^{-e^x-\frac{i\Omega}{g}x}.\label{10-----+++}
\end{equation}

The integral on the right side of \eqref{10-----+++} can be solved by setting $\Omega/g \to z=\Omega/g + i \epsilon$ with $\epsilon \to 0^+$, since
\begin{equation}
   \int_{-\frac{\gamma i \pi}{2} - \infty}^{-\frac{\gamma i \pi}{2} + \infty} dx \, e^{-e^x - i z x} = \Gamma(-i z)\quad, \quad \Im z > 0.
\end{equation}

Therefore, after applying $\epsilon \to 0^+$, we obtain:
\begin{align}
    \alpha^{*}_{\Omega\omega}&=\sqrt{\frac{\Omega}{\omega}}\frac{e^{\frac{\pi\Omega}{2g}}}{2\pi g}\qty(\frac{g}{\omega})^{-\frac{i\Omega}{g}}\Gamma\qty(-\frac{i\Omega}{g})\label{gam+},\\
\beta^{*}_{\Omega\omega}&=-\sqrt{\frac{\Omega}{\omega}}\frac{e^{-\frac{\pi\Omega}{2g}}}{2\pi g}\qty(\frac{g}{\omega})^{-\frac{i\Omega}{g}}\Gamma\qty(-\frac{i\Omega}{g})\label{gam-}.
\end{align}

From \eqref{gam+} and \eqref{gam-} we notice that
\begin{equation}\label{propF}
\alpha^{*}_{\Omega\omega}=-e^{\frac{\Omega\pi}{g}}\beta^{*}_{\Omega\omega}.
\end{equation}

Using \eqref{propF} in the property \eqref{bogopro} of the Bogoliubov-Valatin coefficients,
\begin{align}
\delta(\Omega-\Omega')&=\int_{0}^{\infty}d\omega \qty(\alpha_{\Omega'\omega}\alpha^{*}_{\Omega\omega}-\beta_{\Omega'\omega}\beta^{*}_{\Omega\omega})\\
&=\int_{0}^{\infty}d\omega \qty(e^{\frac{\qty(\Omega'+\Omega)\pi}{g}}\beta_{\Omega'\omega}\beta^{*}_{\Omega\omega}-\beta_{\Omega'\omega}\beta^{*}_{\Omega\omega}),
\end{align}
we arrive to
\begin{equation}\label{deltabb}
\frac{1}{e^{\frac{(\Omega'+\Omega)\pi}{g}}-1}\delta(\Omega-\Omega')=\int_{0}^{\infty}d\omega\beta_{\Omega'\omega}\beta^{*}_{\Omega\omega}.
\end{equation}

Setting $\Omega' = \Omega$, we obtain a divergent integral, as previously mentioned for equation \eqref{NN}. As can be seen from equation \eqref{mirinfo1}, this divergent result arises from the integration over the entire energy spectrum $\omega$ of the scalar field in Minkowski spacetime. However, more significantly from a physical perspective, it is due to the fact that the entire analysis has been conducted using the plane wave field modes $f_\omega$ and $h_{\omega'}$.\\

In addition, the left-hand side of \eqref{deltabb} is a density for a specific value of $\Omega$, which is the energy for the scalar field in Rindler,
\begin{equation}\label{deltabb2}
\mel{0_M}{N_R(\Omega)}{0_M}=\frac{1}{e^{\frac{2\pi\Omega}{g}}-1}\delta(0)=\int_{0}^{\infty}d\omega \abs{\beta_{\Omega\omega}}^{2}.
\end{equation}

The quantity next to $\delta(0)$ is the mean number of particles at energy $\Omega$ for the right modes, in the right wedge of Rindler spacetime, with respect to the vacuum state in Minkowski,
\begin{equation}\label{db-ei}
n_R(\Omega)=\frac{1}{e^{\frac{2\pi\Omega}{g}}-1}.
\end{equation}

Therefore, the accelerated observer (in region I) describes the Minkowski vacuum state using Bose-Einstein statistics, which becomes evident when considering a massless scalar field. In this context, the temperature, given by the inverse of $\beta=2\pi/g$, is proportional to the observer's acceleration:
\begin{equation}\label{215temunru}
T=\beta^{-1}=\frac{g}{2\pi}.
\end{equation}

This is the Unruh effect (also known as the Fulling–Davies–Unruh effect), named after Stephen Fulling, Paul Davies, and W. G. Unruh \cite{fulling1973nonuniqueness,davies1975scalar,Unruh:1976db} (see also \cite{Arzano:2020mhg,Arzano:2020ucu} for a more in-depth analysis from the perspective of group theory).
\subsection{Unruh effect analogue for a Schwarzschild black hole}

The Schwarzschild metric is an exact solution of Einstein’s field equations \cite{schwarzschild1916gravitationsfeld}, which describes the spacetime surrounding a spherical black hole with mass $ M $, without charge and angular momentum. The line element is given by
\begin{equation}
    ds^2 = -\left( 1 - \frac{2M}{r} \right) dt^2 + \left( 1 - \frac{2M}{r} \right)^{-1} dr^2 + r^2 d\Omega^2,\label{schw78}
\end{equation}
where $ d\Omega^2 = d\theta^2 + \sin^2(\theta) d\phi^2 $.\\

As we can observe, the metric is singular at $ r = 2M $. However, this singularity is purely in the coordinates (similar to the Rindler case discussed in \eqref{rind-rho-ome222} for $ \rho \to 0 $). As we will see, the metric near the horizon is conformally equivalent to Minkowski spacetime. The region defined by $ r = 2M $ is known as the event horizon, which represents a causal boundary between the exterior and interior regions of the black hole.\\

According to the equivalence principle, a locally accelerated system is indistinguishable from one immersed in a gravitational field, as perceived by an observer within such systems. Hence, an observer undergoing uniform acceleration in Minkowski spacetime, whose trajectory for a fixed position $ \xi $ is described by equation \eqref{xt2e}, will detect the existence of causal horizons and a thermal bath, a phenomenon referred to as the Unruh effect. This is comparable to the perception of an observer located at a fixed $ r $ in the Schwarzschild metric. Moreover, as discussed in section \ref{caustru}, Schwarzschild and Rindler metrics exhibit quite similar causal structures.\\

To explore the connection between Schwarzschild and Rindler metrics further, let us perform some transformations on the metric \eqref{schw78}. The part corresponding to $ \qty{t, r} $ can be rewritten as
\begin{align}
    d\bar{s}^2 = \left( 1 - \frac{2M}{r} \right) \left( - dt^2 + \frac{dr^2}{\left( 1 - \frac{2M}{r} \right)^2} \right).\label{schwz10}
\end{align}

Let us define the coordinate $ r^*(r) $, known as the Regge-Wheeler tortoise coordinate, through
\begin{equation}
    d{r_*}^2(r) = \frac{dr^2}{\left( 1 - \frac{2M}{r} \right)^2},
\end{equation}
from which we obtain $ r_*(r) = r + 2M \ln \abs{r - 2M} + c $, where $ c $ is an integration constant, fixed as $ c = -2M \ln (2M) $. Thus, we arrive at
\begin{equation}
    r_*(r) = r + 2M \ln \abs{\frac{r}{2M} - 1}.\label{torrrri}
\end{equation}

As we can see, the range of $ r_* $ spans $ -\infty < r_* < \infty $, meaning that the coordinate $ r_* $ maps $ r \to 2M $ to $ r_* \to -\infty $. Additionally, we have
\begin{align}
2r_* &= 2r + 4M \ln \abs{\frac{r}{2M} - 1} \nonumber \\
     &= 2r + 4M \ln \left( \frac{r}{2M} \right) + 4M \ln \abs{1 - \frac{2M}{r}}.\label{v-uschw5}
\end{align}

Solving for the logarithm of the last term, we get
\begin{equation}
    \ln \abs{1 - \frac{2M}{r}} = \frac{2r_* - 2r - 4M \ln \left( \frac{r}{2M} \right)}{4M},
\end{equation}
and by exponentiation on both sides, and knowing that $ r > 2M $, we obtain
\begin{align}
    \left( 1 - \frac{2M}{r} \right) &= e^{\frac{2r_*}{4M}} e^{-\frac{r}{2M}} \frac{2M}{r}.
\end{align}

Thus, the metric \eqref{schwz10} reduces to
\begin{align}
    d\bar{s}^2 = \left( e^{-\frac{r}{2M}} \frac{2M}{r} \right) \left[ e^{\frac{2r_*}{4M}} \left( - dt^2 + dr_*^2 \right) \right],\label{schwz102}
\end{align}
in which we can see that the term in square brackets is analogous to the Rindler line element obtained in \eqref{dsrin}, with $ g = 1/4M $.\\

We define the null coordinates for $ \qty{t, r_*} $ as
\begin{align}
    u = t - r_*\quad,&\quad v = t + r_*,\label{ubarvbarctmr0}\\
    -\infty < u < \infty\quad,&\quad -\infty < v < \infty,
\end{align}
from which the line element \eqref{schwz102}, in terms of the null coordinates $ \qty{u, v} $ becomes
\begin{equation}
    d\bar{s}^2 = \left( e^{-\frac{r}{2M}} \frac{2M}{r} \right) \left[ - e^{\frac{v - u}{4M}} dudv \right],\label{schwz103}
\end{equation}
where the metric remains singular at $ r = 2M $ due to the exponential factor, as $ 2r_* = v - u \to -\infty $ for $ r = 2M $, as seen from \eqref{v-uschw5}.\\

Similarly, we can observe the resemblance between the term in brackets in \eqref{schwz103} and the line element \eqref{e-vmetric}. Thus, in analogy with \eqref{ubar-e-u} and \eqref{vbar-e-v}, we define the map given by the null coordinates $\qty{\bar{u},\bar{v}}$:
\begin{align}
    \bar{u}=-4Me^{-\frac{u}{4M}}\quad&, \quad \bar{v}=4Me^{\frac{v}{4M}},\label{ubarvbarctmr}\\
    -\infty<\bar{u}<0\quad&, \quad0<\bar{v}<\infty.\label{ubarvbarctmr2}
\end{align}

Therefore, the line element \eqref{schwz103} reduces to
\begin{equation}
    d\bar{s}^2=\qty(e^{-\frac{r}{2M}}\frac{2M}{r})(-d\bar{u}d\bar{v}),
\end{equation}
where we notice that the metric is no longer singular at $r=2M$.\\

In this way, as we did in the Rindler analysis, since we have managed to avoid the singularity at $r=2M$, we can extend the range of values of $\qty{\bar{u},\bar{v}}$. To do so, we take the complement of what was seen in \eqref{ubarvbarctmr} and \eqref{ubarvbarctmr2}:
\begin{align}
    \bar{u}=4Me^{-\frac{u}{4M}}\quad&,\quad \bar{v}=-4Me^{\frac{v}{4M}},\label{ubarvbarctmr3}\\
    0<\bar{u}<\infty\quad&,\quad -\infty<\bar{v}<0.\label{ubarvbarctmr4}
\end{align}

Thus, from \eqref{ubarvbarctmr2} and \eqref{ubarvbarctmr4}, we have the extended forms
\begin{equation}
    -\infty<\bar{u}'<\infty\quad,\quad  -\infty<\bar{v}'<\infty.
\end{equation}

This motivates the association of the extended null coordinates $\qty{\bar{u}',\bar{v}'}$ with the coordinates $\qty{T,X}$, such that
\begin{align}
    \bar{u}'=T-X\quad&,\quad\bar{v}'=T+X,\label{ubarvbarctmr5}\\
    -\infty<T<\infty\quad&,\quad  -\infty<X<\infty.
\end{align}

Then, we obtain the line element in the Kruskal-Szekeres (K-S) coordinates \cite{Kruskal:1959vx,Szekeres:1960gm}:
\begin{equation}
    d\bar{s}^2=\qty(e^{-\frac{r}{2M}}\frac{2M }{r})(-dT^2+dX^2).
\end{equation}

From \eqref{ubarvbarctmr5}, \eqref{ubarvbarctmr}, and \eqref{ubarvbarctmr0}, we obtain the coordinate transformation between K-S coordinates $\qty{T,X}$ and Schwarzschild coordinates $\qty{t,r_*}$ for the region $r>2M$:
\begin{align}
    T=4M\exp{\frac{r_>^*}{4M}}\sinh\qty(\frac{t}{4M})\quad,\quad X=4M\exp{\frac{r_>^*}{4M}}\cosh\qty(\frac{t}{4M}),\label{schrind}
\end{align}
where $r_>^*$ is the function given in \eqref{torrrri} for $r>2M$. Thus, we have
\begin{equation}
    \exp{\frac{r_>^*}{4M}}=\qty(\frac{r}{2M}-1)^{1/2}e^{\frac{r}{4M}}.
\end{equation}

Similarly, from \eqref{ubarvbarctmr5}, \eqref{ubarvbarctmr3}, and \eqref{ubarvbarctmr0}, we obtain the transformation for the region $0<r<2M$:
\begin{align}
    T=4M\exp{\frac{r_<^*}{4M}}\cosh\qty(\frac{t}{4M})\quad,\quad X=4M\exp{\frac{r_<^*}{4M}}\sinh\qty(\frac{t}{4M}),
\end{align}
where $r_<^*$ is the function given in \eqref{torrrri} for $r<2M$. Thus, we have
\begin{equation}
    \exp{\frac{r_<^*}{4M}}=\qty(1-\frac{r}{2M})^{1/2}e^{\frac{r}{4M}}.
\end{equation}

As we can observe, the transformation given for the region outside the black hole ($r>2M$) in \eqref{schrind} is very similar to the transformation between Minkowski and Rindler found in \eqref{crde000} and \eqref{crde0001}, where $1/4M$ is analogous to $g$: 
\begin{align}
    T=4M\exp{\frac{r_>^*}{4M}}\sinh\qty(\frac{t}{4M})\quad,\quad t=\frac{e^{g\xi}}{g}\sinh{(g\tau)},\\
    X=4M\exp{\frac{r_>^*}{4M}}\cosh\qty(\frac{t}{4M})\quad,\quad x=\frac{e^{g\xi}}{g}\cosh{(g\tau)}.
\end{align}

\begin{figure}[t]
\centering
     \includegraphics[width=0.4\linewidth]{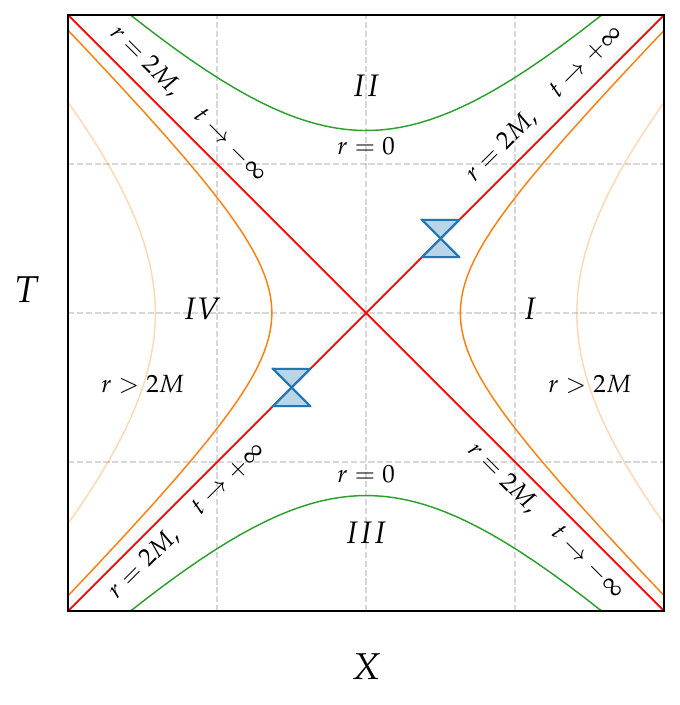}
     \caption{Extended Schwarzschild geometry (restricted to region I) described by the Kruskal-Szekeres coordinates $\qty{T, X}$.}\label{fig13}
\end{figure}

Furthermore, in figure \ref{fig13}, we observe that the Kruskal-Szekeres coordinates, $\qty{T,X}$, provide an extension of the Schwarzschild geometry, which is restricted to the right wedge of the Kruskal-Szekeres diagram (Region I). Additionally, the trajectories of an observer in Schwarzschild spacetime at a fixed $r_*$ are represented by the orange semi-hyperbolas in region I (figure \ref{fig13}), described by the equation:
\begin{equation}
    X^2 - T^2 = (4M)^2 \exp\left(\frac{2r_*}{4M}\right).
\end{equation}

This expression is, of course, analogous to the one obtained for Rindler in \eqref{xt2e}. Additionally, we know that the temperature associated with a black hole is given by $T=\kappa/2\pi$, where $\kappa$ is the surface gravity. Therefore, the analogy between Schwarzschild and Rindler suggests that 
\begin{equation}
    T:\quad \frac{g}{2\pi} \to \frac{\kappa}{2\pi},
\end{equation}
and precisely, the value of the surface gravity for Schwarzschild is $\kappa=1/4M$. Hence, the temperature is correct, which validates the analogy with the Unruh effect:
\begin{equation}
    T=\frac{1}{8\pi M}.
\end{equation}

\subsection{Particle detector}\label{sec.3va-or}
In section \ref{tervcuuaob}, we have performed the calculation that demonstrates the Unruh effect based on the analysis of the expected value of the number operator (see equations \eqref{deltabb2} - \eqref{215temunru}). However, for experimental purposes, the number operator is not a suitable observable if we consider a system with a large number of particles (statistical ensemble). Therefore, in this section, we will examine the simplest observable, given by the response of a moving detector coupled to a massless real scalar field in Minkowski spacetime.\\

The device we will analyze is known in the literature as the Unruh-DeWitt detector (see \cite{Unruh:1976db} and \cite{dewitt1979quantum}), which moves along the worldline described by $x^\mu(\tau)$, where $\tau$ is the proper time of the detector. The interaction of this detector with the massless real scalar field $\phi(x^\mu)$ is described in the Lagrangian by $\mathcal{L}_{I} = \alpha m(\tau) \phi(x^\mu(\tau))$, where $m(\tau)$ is the detector's monopole moment operator and $\alpha$ is the coupling constant.\\

The initial state of the field-detector system is described by $\ket{0_M, E_0} \equiv \ket{0_M} \otimes \ket{E_0}$, where $\ket{0_M}$ represents the vacuum state of the field defined in Minkowski geometry, and $\ket{E_0}$ denotes the ground state of the detector. Let us consider that during its trajectory, the detector transitions to the excited state $\ket{E}$, while the field transitions to the excited state $\ket{\psi}$. Then, the transition amplitude between the initial and final states is given by
\begin{equation}
    T_{ex}=\bra{E,\psi}T\exp\qty[i\int_{-\infty}^\infty d\tau \mathcal{L}_I(\tau)]\ket{0_M,E_0},\label{??..,,,+++}
\end{equation}
where $T$ is the time-ordering operator. Likewise, assuming that the coupling constant $\alpha$ is sufficiently small so that radiative corrections to the interaction Lagrangian can be neglected, the amplitude \eqref{??..,,,+++} can be adequately approximated by the first-order term in perturbation theory (for further details, see \cite{dewitt1979quantum, birrell1984quantum}):  
\begin{equation}
    T_{ex}\approx i\alpha\bra{E,\psi}\int_{-\infty}^\infty d\tau\, m(\tau)^{(IP)} \phi(x^\mu)^{(IP)}\ket{0_M,E_0},
\end{equation}
where $(IP)$ denotes that the field is in the interaction picture\footnote{Recall that the relationship between the Schrödinger, Heisenberg, and interaction pictures is given by  
\begin{equation}  
    \mathcal{O}^{(IP)}(t) = e^{iH_0^{(S)}t} \mathcal{O}^{(S)} e^{-iH_0^{(S)}t} = e^{iH_0^{(S)}t} e^{-iHt} \mathcal{O}^{(H)}(t) e^{iHt} e^{-iH_0^{(S)}t},  
\end{equation}  
where $H = H_0 + H_I$ is the total Hamiltonian in the Heisenberg picture.}. Furthermore, since $ m(\tau)^{(IP)} = e^{iH_0 \tau} m(0) e^{-iH_0 \tau} $, the transition amplitude is given by
\begin{eqnarray}  
    T_{ex} &=& i\alpha\int_{-\infty}^{\infty} d\tau \bra{E}e^{iH_0 \tau} m(0) e^{-iH_0 \tau} \ket{E_0} \bra{\psi}\phi(x^\mu)\ket{0_M} \nonumber\\  
    &=& i\alpha\int_{-\infty}^{\infty} d\tau \bra{E}m(0)\ket{E_0} e^{i(E-E_0)\tau} \bra{\psi}\phi(x^\mu)\ket{0_M},\label{psiphisuu0m} 
\end{eqnarray}  
where for simplicity, the use of the superscript $(IP)$ in the field operator has been omitted.\\

Then, the transition probability for all $E$ and $\psi$ will be given by the following summations:
\begin{eqnarray}
    P_{ex}&=&\sum_E\sum_\psi \abs{T_{ex}}^2\nonumber\\
    &=&\alpha^2 \int_{-\infty}^{\infty} \int_{-\infty}^{\infty} d\tau d\tau^\prime \sum_E\abs{\bra{E}m(0)\ket{E_0}}^2 e^{-i(E-E_0)(\tau-\tau^\prime)}\times\nonumber\\
    &&\hspace{6cm}\times\bra{0_M} \phi(x^\mu(\tau)) \phi(x^\mu(\tau^\prime)) \ket{0_M},\label{probexxx}
\end{eqnarray}
where we have used $\sum_\psi \ket{\psi}\bra{\psi} = 1$.\\

Thus, as we can observe so far, the transition probability given by
\begin{equation}  
    P_{ex}=\underbrace{\alpha^2  \sum_E\abs{\bra{E}m(0)\ket{E_0}}^2}_{(2)}\underbrace{\int_{-\infty}^{\infty} \int_{-\infty}^{\infty} d\tau d\tau^\prime e^{-i(E-E_0)(\tau-\tau^\prime)} G^+(t,\mathbf{x};t^\prime,\mathbf{x}^\prime)}_{(1)}\label{probxxxxex}
\end{equation}
depends on two components: (1) the detector response function, which describes the bath of “particles” that the detector effectively experiences as a result of its motion; and (2) the selectivity of the detector with respect to the bath, which depends on its internal structure. Furthermore, we have identified the  Wightman function:
\begin{equation}
    G^+(t,\mathbf{x};t^\prime,\mathbf{x}^\prime)=\bra{0_M} \phi(t,\mathbf{x}) \phi(t^\prime,\mathbf{x}^\prime) \ket{0_M}.
\end{equation}

For the calculation of the probability in \eqref{probxxxxex}, we will develop the computation of the Wightman function, for the massless real scalar field in 4 dimensions, as it is a well-known result in the literature (see, for example, \cite{birrell1984quantum}), and whose final conclusion will coincide with that obtained in \eqref{db-ei}. Then, we have (see \cite{weinberg1995quantum,srednicki2007quantum}):
\begin{equation}
    G^+(t,\mathbf{x};t^\prime,\mathbf{x}^\prime) = \int \frac{d^3k}{(2\pi)^3} \frac{e^{ik^\mu (x_\mu - x^\prime_\mu)}}{2 \omega_{\mathbf{k}}},\label{wenynyny}
\end{equation}
where, in our specific case, $k^0 = \omega_\mathbf{k} = |\mathbf{k}|$, since we are analyzing a massless field. In addition, it is important to mention that in the Wightman function $(t-t')\in\mathbb{R}$.\\

Let us express the integral in \eqref{wenynyny}, in spherical coordinates:
\begin{eqnarray}
    G^+(t,\mathbf{x};t^\prime,\mathbf{x}^\prime)&=&\frac{1}{(2\pi)^3}\int_0^{2\pi}d\varphi \int_0^\infty dk k^2  \frac{e^{-ik(t-t^\prime)}}{2k}\int_0^\pi d\theta\sin(\theta) e^{ik\abs{\mathbf{x}-\mathbf{x}^\prime}\cos(\theta)}\nonumber\\
   &=&\frac{2}{(2\pi)^2\abs{\mathbf{x}-\mathbf{x}^\prime}}\int_0^\infty dk k \frac{e^{-ik(t-t^\prime)}}{2k}\sin\qty(k\abs{\mathbf{x}-\mathbf{x}^\prime})\label{divdiv7&&**}.
\end{eqnarray}

Then, in order to perform the integral over $k$, it is necessary to carry out the analytic continuation in the time interval through $(t - t') \to (t - t' - i\epsilon)$, where the infinitesimal parameter $\epsilon \to 0^+$ ensures convergence. In this way, we obtain:
\begin{eqnarray}
    G^+(t,\mathbf{x};t^\prime,\mathbf{x}^\prime)
    &\approx& -\frac{1}{4\pi^2}\qty[\qty(t-t^\prime - i\epsilon)^2 - \abs{\mathbf{x}-\mathbf{x}^\prime}^2]^{-1}. \label{g+txtpcp}
\end{eqnarray}

Let us now consider that the detector follows an inertial trajectory given by  
\begin{equation}
    \mathbf{x} = \mathbf{x}_0 + \mathbf{v}t, \quad t = \tau(1 - \abs{\mathbf{v}}^2)^{-1/2},
\end{equation} 
where we can set $\mathbf{x}_0 = 0$. Then, in \eqref{g+txtpcp}, we have:  
\begin{eqnarray}
G^+(t,\mathbf{x};t^\prime,\mathbf{x}^\prime)_{\text{inertial}} &=& -\frac{1}{4\pi^2}\frac{1}{\qty[(\tau-\tau^\prime)(1-\abs{\mathbf{v}}^2)^{-1/2} - i\epsilon]^2 - \abs{\mathbf{v}}^2(\tau-\tau^\prime)^2(1-\abs{\mathbf{v}}^2)^{-1}}\nonumber\\
&=&-\frac{1}{4\pi^2}\frac{1}{(\tau-\tau^\prime)^2 - 2i\epsilon(\tau-\tau^\prime)(1-\abs{\mathbf{v}}^2)^{-1/2} - \epsilon^2}\nonumber\\
&=&-\frac{1}{4\pi^2}\frac{1}{(\tau-\tau^\prime - i\epsilon)^2},\label{ghyninertiallll}
\end{eqnarray}  
where the positive factor $(1 - |\mathbf{v}|^2)^{-1/2}$ has been absorbed into $\epsilon$, which evidently modifies the units of the `new' infinitesimal parameter $\epsilon$. It is therefore important to take this detail into account.\\

On the other hand, considering that the detector follows a non-inertial trajectory characterized by constant acceleration and described by $x = \sqrt{t^2 + 1/g^2}$ and $y = z = 0$, with $g$ constant, as derived from \eqref{rindler1} and \eqref{rindler2}, we have: 
\begin{eqnarray}
    G^+(t,x;t^\prime,x^\prime)_{\text{non-inertial}} &=& -\frac{1}{4\pi^2} \left[-\frac{2}{g^2} - \epsilon^2 + \frac{2}{g^2} \cosh\qty(g(\tau - \tau^\prime)) \right.\nonumber\\
    &&\hspace{2cm}\left. -\,\,i\epsilon \qty(\frac{2}{g} \sinh(g\tau) - \frac{2}{g} \sinh(g\tau^\prime))\right]^{-1}.
\end{eqnarray}

Knowing that the infinitesimal parameter $\epsilon \to 0^+$ can absorb any positive factor, we can implement the following replacement:
\begin{equation}
i\epsilon \qty[\frac{2}{g} \sinh(g\tau) - \frac{2}{g} \sinh(g\tau^\prime)] \;\to\; (i\epsilon)\,\text{sgn}(\tau-\tau'),
\end{equation}
where $\text{sgn}(\tau-\tau')$ denotes the sign function. In this way, we obtain:
\begin{equation}
    G^+(t,x;t^\prime,x^\prime)_{\text{non-inertial}}\approx -\frac{1}{4\pi^2} \qty[-\frac{2}{g^2} + \frac{2}{g^2} \cosh\qty(g(\tau - \tau^\prime)) - (i\epsilon)\text{sgn}(\tau-\tau')]^{-1}.\label{giogo}
\end{equation}

In addition, we can note that the factor in brackets in \eqref{giogo} matches the following function:
\begin{eqnarray}
    \frac{4}{g^2}\sinh^2\qty[\frac{g(\tau - \tau^\prime)}{2} - i\epsilon]&=&-\frac{2}{g^2} + \frac{2}{g^2}\cosh\left[g(\tau - \tau^\prime) - 2i\epsilon\right]\nonumber\\
    &=&-\frac{2}{g^2} + \frac{2}{g^2} \cos(\epsilon)^2 \cosh\left[g (\tau - \tau^\prime)\right]+\nonumber\\
    &&- \frac{2}{g^2} \cosh\left[g (\tau - \tau^\prime)\right] \sin(\epsilon)^2+ \nonumber\\
    &&- \frac{4i}{g^2} \cos(\epsilon) \sin(\epsilon) \sinh\left[g (\tau - \tau^\prime)\right]\nonumber\\
    &\approx&-\frac{2}{g^2} + \frac{2}{g^2}\cosh\left[g (\tau - \tau^\prime)\right]-(i\epsilon)\text{sgn}(\tau-\tau'),
\end{eqnarray}
where we have used the approximations $\cos^2(\epsilon) \approx 1$, 
$\sin^2(\epsilon) \approx \epsilon^2\approx 0$, and $\cos(\epsilon)\sin(\epsilon) \approx \epsilon$, 
together with the replacement
\begin{equation}
\frac{4i}{g^{2}}\,\epsilon\,\sinh\!\big[g(\tau-\tau')\big]\to(i\epsilon)\,\text{sgn}(\tau-\tau').
\end{equation}

Thus, we find that $G^+(t,x;t^\prime,x^\prime)_{\text{non-inertial}} = -g^2 / \qty[16\pi^2 \sinh^2\qty(g(\tau - \tau^\prime) / 2 - i\epsilon)]$. In addition, since $1 / \sinh^2(\pi x) = \sum_{n=-\infty}^\infty \pi^{-2} (x + in)^{-2}$, we have
\begin{eqnarray}
    G^+(t,x;t^\prime,x^\prime)_{\text{non-inertial}}&=&-\frac{g^2}{16\pi^2}\frac{1}{\sinh^2\qty[\frac{g(\tau - \tau^\prime)}{2} - i\epsilon]}\nonumber\\
    &=&-\frac{g^2}{16\pi^2}\sum_{n=-\infty}^\infty\pi^{-2}\qty(\frac{g(\tau-\tau^\prime)}{2\pi} - \frac{i\epsilon}{\pi}+in)^{-2}.\label{g+++exxx}
\end{eqnarray}

As we can observe, the Wightman Green function, both for the inertial case \eqref{ghyninertiallll} and the non-inertial case \eqref{g+++exxx}, depends on $\tau - \tau^\prime$. This implies that the integrand in the response function \eqref{probxxxxex} is invariant under proper-time translations: $\tau, \tau^\prime \to \tau, \tau^\prime + \text{constant}$. Consequently, the detector-field system is in a stationary state from the detector's perspective. Furthermore, by performing the variable change $\tau_\pm = \tau \pm \tau^\prime$, it becomes evident that the integral in \eqref{probxxxxex} may diverge, as it is given by the product of the Fourier transform of $G^+(\tau_-)$ and an infinite integral over $\tau_+$:
\begin{equation}
    \frac{1}{2}\int_{-\infty}^\infty d\tau_+ \int_{-\infty}^\infty d\tau_- \, e^{-i(E - E_0)\tau_-} G^+(\tau_-).\label{divdivg&&}
\end{equation}  
For this reason, instead of calculating the transition probability, we will compute the transition rate $R_{ex}$:
\begin{equation}
   R_{ex}=\frac{d P_{ex}}{d\tau_+}=\frac{\alpha^2}{2}  \sum_E\abs{\bra{E}m(0)\ket{E_0}}^2 \int_{-\infty}^{\infty} d\tau_- e^{-i(E-E_0)\tau_-} G^+(\tau_-).\label{372exxme}
\end{equation}

\begin{figure}[t]
\centering
     \includegraphics[width=0.4\linewidth]{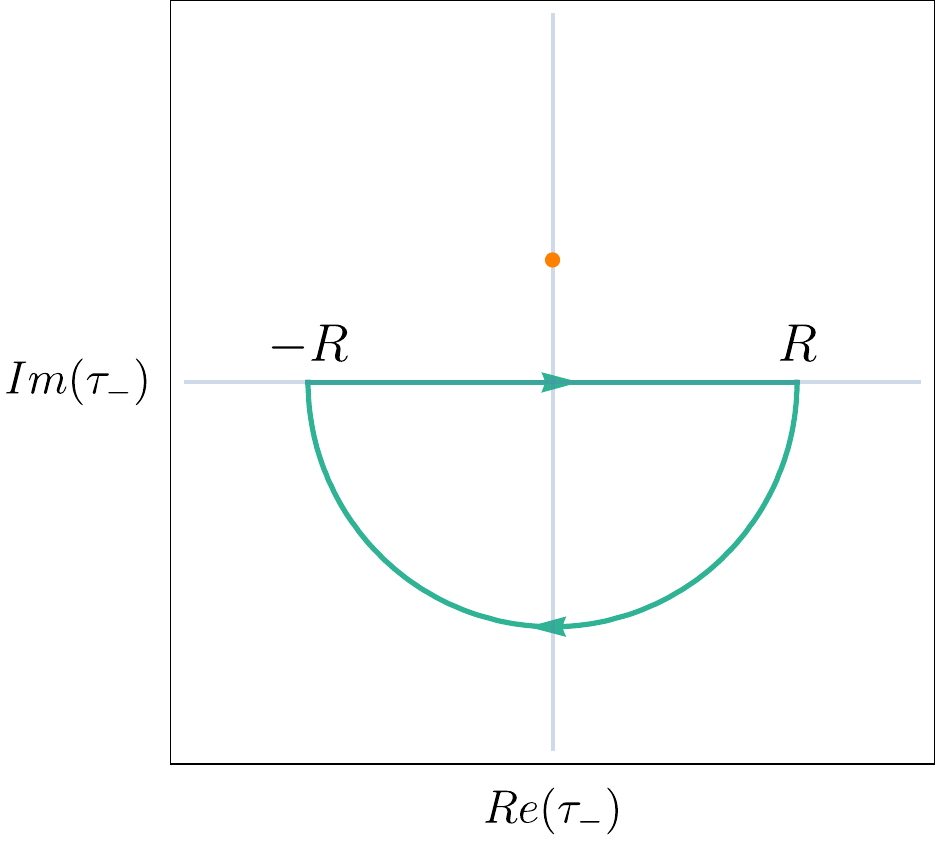}
     \caption{Contour in the complex $\tau_-$-plane.}\label{h1???*}
\end{figure}

For the detector following an inertial trajectory, from \eqref{ghyninertiallll} and \eqref{372exxme}, we obtain  
\begin{equation}
    R_{ex}^{\text{inertial}}=-\frac{\alpha^2}{8\pi^2}  \sum_E\abs{\bra{E}m(0)\ket{E_0}}^2 \int_{-\infty}^{\infty} d\tau_- \, e^{-i(E-E_0)\tau_-} \frac{1}{(\tau_- - i\epsilon)^2}.
\end{equation} 
Since $E - E_0 > 0$, the above integral in the complex $\tau_-$-plane is evaluated using the contour shown in figure \ref{h1???*}, which excludes the second-order pole at $\tau_-=i\epsilon$. Therefore, the integral vanishes. This result is consistent, as an inertial observer detects no discrepancy in the physics of the vacuum state $\ket{0_M}$. Therefore, the transition probability is zero (since the integral over $\tau_-$ in \eqref{divdivg&&} vanishes):
\begin{equation}
    P_{ex}^{\text{inertial}} = 0.\label{probzzzz00}
\end{equation}

On the other hand, for the non-inertial observer, from \eqref{g+++exxx} and \eqref{372exxme}, we have
\begin{equation}
    R_{ex}^{\text{non-inertial}}=-\frac{\alpha^2}{8\pi^2}\sum_E\abs{\bra{E}m(0)\ket{E_0}}^2\sum_{n=-\infty}^\infty\int_{-\infty}^{\infty} d\tau_-\frac{e^{-i(E-E_0)\tau_-}}{\qty(\tau_- -\frac{2i\epsilon}{g}+\frac{2\pi in}{g})^2},\label{81..,,20-}
\end{equation}
where we can observe a second-order pole at $\tau_- = \frac{2i}{g}(\epsilon - \pi n)$. In this way, for $n \leq 0$, the pole is located in the upper part of the complex $\tau_-$-plane, while for $n > 0$, the pole is located in the lower part.\\

\begin{figure}[t]
\centering
     \includegraphics[width=0.4\linewidth]{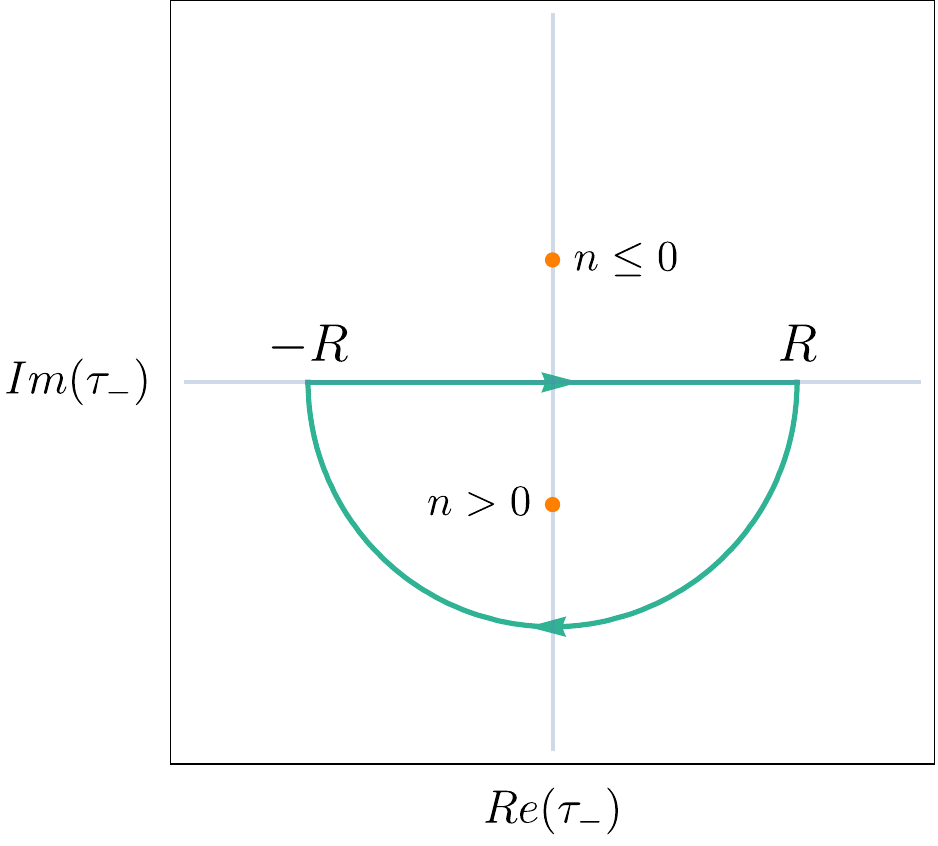}
     \caption{Contour in the complex $\tau_-$-plane.}\label{h1???}
\end{figure}

Thus, since $E - E_0 > 0$, we will follow the contour shown in figure \ref{h1???}, which implies that $n > 0$. Therefore, the summation over $n$ will start at $n = 1$:
\begin{eqnarray}
    \sum_{n=-\infty}^\infty\int_{-\infty}^{\infty} d\tau_-\frac{e^{-i(E-E_0)\tau_-}}{\qty(\tau_- -\frac{2i\epsilon}{g}+\frac{2\pi in}{g})^2}&=&\sum_{n=1}^\infty(-2\pi i)(-i(E-E_0))e^{-i(E-E_0)(\frac{2i\epsilon}{g}-\frac{2\pi in}{g})}\nonumber\\
    &=&-2\pi(E-E_0)e^{\frac{2\epsilon}{g}(E-E_0)}\sum_{n=1}^\infty e^{-\frac{2\pi n}{g}(E-E_0)}\nonumber\\
    &=&-2\pi(E-E_0)e^{\frac{2\epsilon}{g}(E-E_0)}\frac{1}{e^{\frac{2\pi}{g}(E-E_0)}-1}.
\end{eqnarray}

In this way, the transition rate \eqref{81..,,20-} (taking $\epsilon\to 0^+$) for the non-inertial observer is given by:
\begin{equation}
    R_{ex}^{\text{non-inertial}}=\frac{\alpha^2}{4\pi}  \sum_E(E-E_0) \abs{\bra{E}m(0)\ket{E_0}}^2\frac{1}{e^{\frac{2\pi}{g}(E-E_0)}-1}.\label{plnnnnnffsscc2}
\end{equation}

As we can observe, the transition rate is characterized by the Planck factor
\begin{equation}
    \frac{1}{e^{\frac{2\pi}{g}(E-E_0)}-1},
\end{equation}
which indicates that the excitation detected by the Unruh-DeWitt detector, under constant acceleration, occurs precisely because, from its perspective, the Minkowski vacuum behaves as a thermal bath described by the Bose-Einstein distribution at a temperature $T=g/2\pi$. This is the Unruh effect.\\

Furthermore, we can observe from \eqref{probzzzz00} and \eqref{plnnnnnffsscc2} that it is the constant acceleration of the detector, induced by an external force $F_{\text{ext}}$, that generates the emission of quanta from the field along the trajectory of the device. This occurs due to the coupling between the field and the detector. Moreover, this emission of quanta produces a resistance against the force responsible for the acceleration. Therefore, the work performed by this external force to overcome the resistance is such that it provides the energy required for the excitation of the field (and consequently the emission of quanta and generation of resistance) and the detector: $\ket{0_M, E_0}\to\ket{\psi, E}$. 

\section{Euclidean approach and the thermofield double state}\label{sec4...}
\subsection{Euclidean path integral}\label{epathinsec}

The transition amplitude due to the time evolution operator $e^{-iHt}$, given by
\begin{equation}
\braket{\phi(x_f,t_f=t)}{\phi(x_\circ,t_\circ=0)}=\mel{\phi_f}{e^{-iHt}}{\phi_\circ},
\end{equation}
is expressed as a Feynman Path Integral \cite{feynman2010quantum}:
\begin{equation}\label{22amp1}
\mel{\phi_f}{e^{-iHt}}{\phi_\circ}=\displaystyle\int_{\phi(t_\circ)=\phi_\circ}^{\phi(t)=\phi_f} D\qty[\phi] e^{iI\qty[\phi]},
\end{equation}
where the integration limits correspond to the initial and final states, and $I$ is the classical action of the theory, $I\qty[\phi]=\int dt dx\mathcal{L}(\phi).$\\

Let us develop the above analysis in the \textit{Euclidean signature}. To do that, we apply the Wick rotation\footnote{It can be interpreted as a clockwise rotation of $\pi/2$ in the complex $t$-plane. Consider the analytic continuation of $t$, given by $z_t = t + is$. Initially, $z_t$ lies on the real axis; then, after applying the aforementioned rotation, it will be located on the negative part of the imaginary axis, resulting in the transformation $t \to -is$, for $s > 0$. Similarly, by relabeling $s$ as $t_E$, the final effect of this transformation can be understood through the substitution $t = -it_E$, which corresponds precisely to the Wick rotation mentioned in \eqref{221wick}.}
\begin{equation}
t_E=it.\label{221wick}
\end{equation}
 
Applying \eqref{221wick} in $i I$, for a real scalar field $\phi$, we obtain
\begin{align}
    i I\qty[\phi]&= i\int dt dx \qty[\frac{1}{2}\qty(\frac{\partial \phi}{\partial t})^2-\frac{1}{2}\qty(\nabla\phi)^2-\frac{1}{2}m^2\phi^2]\nonumber\\&=-\int dt_E dx\qty[\frac{1}{2}\qty(\frac{\partial \phi}{\partial t_E})^2+\frac{1}{2}\qty(\nabla\phi)^2+\frac{1}{2}m^2\phi^2].
\end{align}

Let us define the Euclidean action as
\begin{align}\label{sqbr}
I_E\qty[\phi]:=\int dt_E dx\qty[\frac{1}{2}\qty(\frac{\partial \phi}{\partial t_E})^2+\frac{1}{2}\qty(\nabla\phi)^2+\frac{1}{2}m^2\phi^2].
\end{align}

As we note, the term in square brackets in \eqref{sqbr} is the Hamiltonian density $\mathcal{H}$ with respect to the \textit{Euclidean time} $t_E$. Considering that $\mathcal{H}$ is time-independent, then
\begin{equation}\label{-ie=ht}
i I\qty[\phi] = - I_E\qty[\phi] = -HT,
\end{equation}
where
\begin{align}
H&=\int d\sigma\qty[\frac{1}{2}\qty(\frac{\partial \phi}{\partial t_E})^2+\frac{1}{2}\qty(\nabla\phi)^2+\frac{1}{2}m^2\phi^2],\\
T&=\int dt_E.
\end{align}

Applying the Wick rotation \eqref{221wick} in the amplitude \eqref{22amp1} we obtain the \textit{Euclidean path integral}\footnote{Hereafter, we will refer to it simply as the path integral instead of the Euclidean path integral.}
\begin{equation}\label{22amp2}
\mel{\phi_f}{e^{-Ht_E}}{\phi_\circ}=\displaystyle\int_{\phi(t_E=0)=\phi_\circ}^{\phi(t_E)=\phi_f} D\qty[\phi] e^{-I_E\qty[\phi]},
\end{equation}
where the transition amplitude is given by the operator $ e^{-Ht_E} $.\\

In addition, the state $\ket{\phi}$, propagated from $t_E=-T$ to $t_E=t=0$,
\begin{equation}\label{icesintop}
\ket{\phi(t=0)}=e^{-HT}\ket{\phi(t_E=-T)},
\end{equation}
is given by the path integral without a defined upper limit:
\begin{equation}
\ket{\phi(t=0)}=e^{-HT}\ket{\phi(t_E=-T)}=\displaystyle\int_{\phi(t_E=-T)=\phi_\circ}^{\phi(t=0)}D\qty[\phi] e^{-I_E\qty[\phi]}.
\end{equation}

In general, an arbitrary state at Lorentzian time $t$ can be seen as a result of propagation in two stages: from $t_E=-T$ to $t=t_E=0$, then its propagation in Lorentzian time:
\begin{equation}\label{st243}
\ket{\phi(t)}=e^{-iHt}\qty[e^{-HT}\ket{\phi(t_E=-T)}].
\end{equation}

\subsubsection{Ground state}\label{222estadodevacio}

Let us propagate, by a long Euclidean time $T\rightarrow \infty$, an arbitrary state $\ket{\phi}$ in the basis of the Hamiltonian,
\begin{align}
    e^{-HT}\ket{\phi}&=\displaystyle\sum_n e^{-HT}\ket{n}\bra{n}\ket{\phi}\nonumber\\
&=\displaystyle\sum_n \bra{n}\ket{\phi}e^{-E_n T}\ket{n}.
\end{align}

As the eigenvalue $E_n>0$ increases with $n$, the main contribution to the sum is given by the ground state ($E_0< E_n,\,\forall\,n\geq 1$),
\begin{align}
e^{-HT}\ket{\phi}\approx \bra{0}\ket{\phi}e^{-E_0 T}\ket{0}\label{223need}.
\end{align}

In order to prepare the ground state for its evolution in Lorentzian time, we define $\ket{\phi}$ at $t_E=-T$. Then, the action of $e^{-HT}$ over $\ket{\phi}$ gives us a state at $t=0$,
\begin{align}\label{teusare}
\ket{0} \approx \frac{1}{\bra{0}\ket{\phi}e^{-E_0 T}}e^{-HT}\ket{\phi(t_E=-T)}.
\end{align}

As $T\rightarrow \infty$ corresponds to $t_E\rightarrow -\infty$, we define the ground state, at $t=0$, as the following path integral without the upper limit:
\begin{align}
\ket{0} &\propto \int_{\phi(t_E\rightarrow-\infty)=\phi_\circ}^{\phi(t=0)}D\qty[\phi] e^{-I_E\qty[\phi]}.
\end{align}

\subsubsection{Partition function}\label{222parfun}
The partition function $Z(\beta)$ of a canonical ensemble,
\begin{equation}\label{patichon}
Z(\beta)=\Tr e^{-\beta H}=\sum_n e^{-\beta E_n},
\end{equation}
where $\beta$ is the inverse of temperature, can be written in terms of the Euclidean path integral:
\begin{equation}
    Z(\beta)=\sum_{n}\mel{\phi_{n}}{e^{-\beta H}}{\phi_{n}}=\sum_{n}\oint_{\phi_n}D\qty[\phi] e^{-I_E\qty[\phi]}.
\end{equation}

As the initial and final states are the same, we can infer the periodicity in Euclidean time given by $\beta$,
\begin{equation}
    t_E \sim t_E + \beta\label{tete+bet}.
\end{equation}

Hence,
\begin{equation}
\ket{\phi_n(t_E)}=\ket{\phi_n(t_E+\beta)}.
\end{equation}

\subsubsection{Conical singularity and the associated temperature}\label{unef}
Applying the Wick rotation to the Rindler metric \eqref{rind-rho-ome}, in the form
\begin{equation}
    \tau_E=i\tau,\label{rtwpo}
\end{equation}
we obtain
\begin{equation}
ds^2=\rho^2 g^2d\tau_E^2+d\rho^2.\label{ds2rineuh2}
\end{equation}

As we can see, by setting $\theta = g\tau_E$ in the line element \eqref{ds2rineuh2}, we obtain the polar plane line element
\begin{equation}
    ds^2 = \rho^2 d\theta^2 + d\rho^2.\label{ds2rineuh}
\end{equation}

In order to avoid the conical singularity at the origin of the polar plane, we impose the following periodicity in $\theta$:
\begin{equation}
\theta \sim \theta + 2\pi\label{tetateta2pi}.
\end{equation}

\begin{figure}[t]
\centering
     \includegraphics[width=0.4\linewidth]{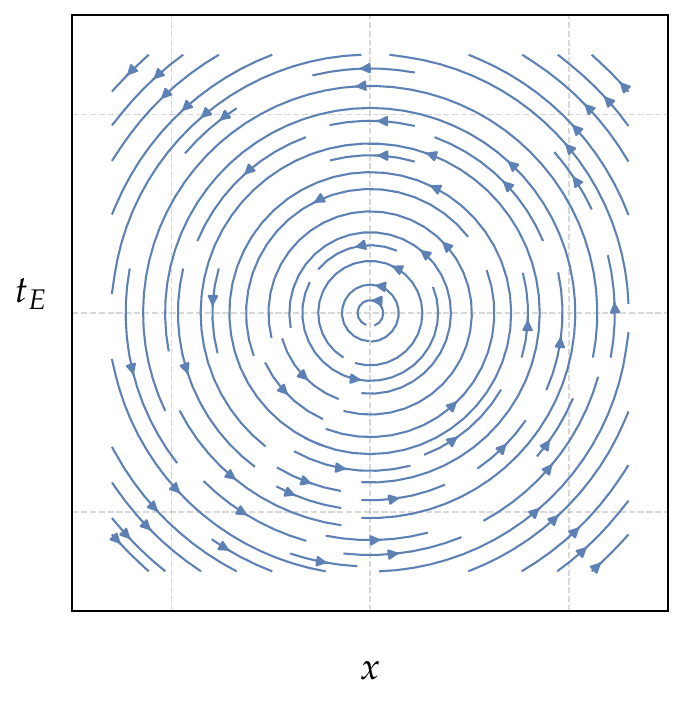}
     \caption{Integral curves of the vector field $\partial_{\tau_E}/g$ in the Euclidean plane.}\label{fig6a}
\end{figure}

Such a condition imposes a periodicity in $\tau_E$. Hence, the periodicity in $ t_E $ is given by
\begin{equation}
\tau_E \sim \tau_E + \frac{2\pi}{g}.\label{tete2pig}
\end{equation}

Therefore, from the result obtained in section \ref{EfectoUnsubse}, we find that a field theory on Euclidean space with a periodicity of $\beta$ in Euclidean time,
\begin{equation}
    \beta=\frac{2\pi}{g}\label{be2piovg},
\end{equation}
can be related to a thermal field theory on Minkowski spacetime at the temperature
\begin{equation}
T=\beta^{-1}=\frac{g}{2\pi},
\end{equation}
which is precisely the Unruh temperature \eqref{215temunru}.

\subsection{Thermofield double state and density matrix}

\subsubsection{Euclidean section and polar coordinates}
Analyzing the geometry from a Euclidean perspective through the application of the Wick rotation \eqref{rtwpo} on the Rindler line element \eqref{rind-rho-ome} and the coordinate transformations \eqref{x-rho-o}, \eqref{x-rho-o4}, and their respective forms for the left wedge (region IV, as mentioned in \eqref{k10-/.}), we have:
\begin{align}
    t_E=\pm\rho\sin\qty(g\tau_E)\label{216po1},\\
    x=\pm\rho\cos\qty(g\tau_E)\label{216po2},
\end{align}
where $\pm$ indicates that this is the Euclidean transformation obtained from regions I and IV of Minkowski, respectively. Moreover, just as in the Lorentzian version of Rindler, in both cases we have the same line element \eqref{ds2rineuh2}.\\

Furthermore, we have that the Killing vector associated with the line element \eqref{ds2rineuh2} is the generator of counterclockwise rotations in the Euclidean plane $\{t_E, x\}$ (see figure \ref{fig6a}), scaled by the factor $g$:
\begin{equation}\label{dthetadrot}
\partial_{\tau_E} = g(x\partial_{t_E} - t_E\partial_x).
\end{equation}

Therefore, in polar coordinates, given by $\theta=g\tau_E$, this operator is precisely the generator of rotations. This makes sense since the rotational periodicity in the plane is $2\pi$. Then, given the relationship between $\tau_E$ and $\theta$, we have
\begin{equation}
    \partial_{\tau_E}=g\partial_\theta.
\end{equation}

Thus, in Euclidean signature, the Hamiltonian operator responsible for generating evolution in $\tau_E$, denoted by $H_R$ (the index $R$ stands for Rindler), will be related to the one given for $\theta$, denoted by $H_\theta$, by
\begin{equation}
    H_R = gH_\theta.\label{ghhtheta}
\end{equation}

\begin{figure}[t]
    \centering
    \begin{subfigure}{0.4\textwidth}
        \includegraphics[width=\linewidth]{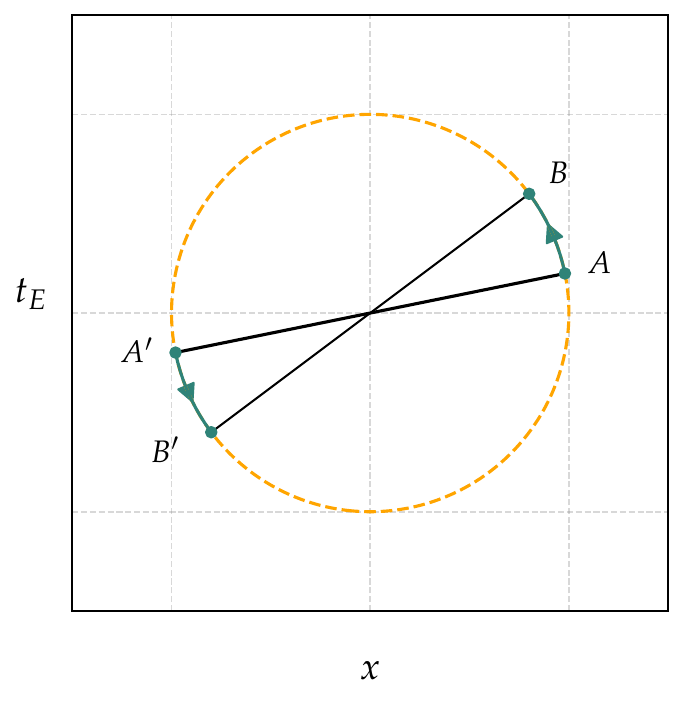}
\captionsetup{justification=centering}
        \caption{}
        \label{fig6b}
    \end{subfigure} 
\hspace{.5cm}
\begin{subfigure}{0.4\textwidth}
        \includegraphics[width=\linewidth]{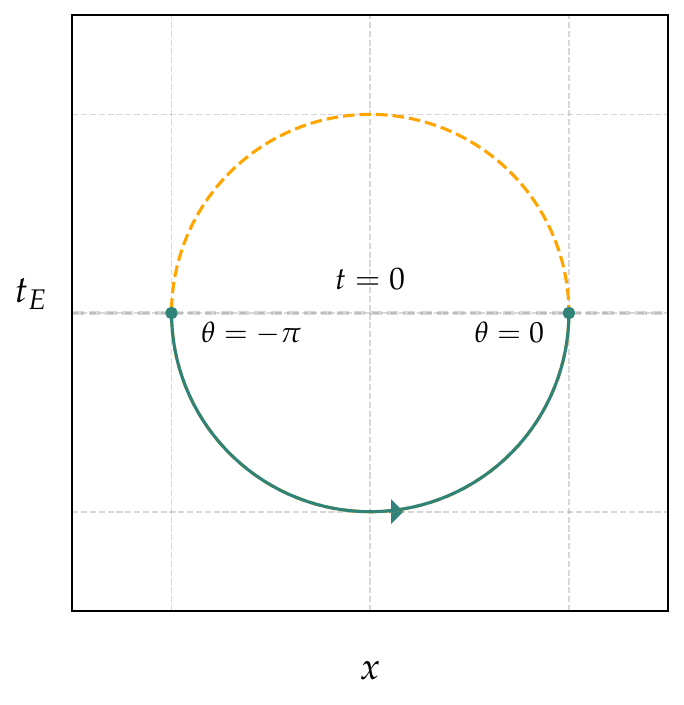}
\captionsetup{justification=centering}
        \caption{}
        \label{fig6c}
    \end{subfigure}
    \caption{(a) Evolution from $A$ to $B$ related by parity and time-reversal to that from $A'$ to $B'$. (b) Evolution by $\pi$ radians in the lower half of the Euclidean plane.}
    \label{fig15}
\end{figure}

It is important to mention that unlike the Lorentzian case, here the transformations do not complement each other in the previously seen sense, where both semi-hyperbolas joined to form a single hyperbola in Minkowski spacetime. Here, in both cases $(\pm)$ of the transformations \eqref{216po1} and \eqref{216po2}, for the entire range of $\theta=g\tau_E$ values, we obtain a circumference in the Euclidean plane, which means that this mapping is 2 to 1. However, complementarity is observed when analyzing $\theta$ evolution, as one trajectory immediately defines the other, again through parity and time-reversal transformations, which in the Euclidean version translates as parity in $t_E$. This is shown in terms of $\theta$ in figure \ref{fig6b}.

\subsubsection{Transition amplitude}\label{transampli}

The transition amplitude between an eigenstate of the field operator $\phi$ in Minkowski space, $\ket{\phi}$, and the ground state $\ket{0}$, at $t = 0$, is given by
\begin{align}
\bra{\phi}\ket{0}&\propto\lim_{T\to\infty}\bra{\phi}e^{-HT}\ket{\phi (t_E=-T)}=\int_{\phi(t_E\rightarrow-\infty)=\phi_\circ}^{\phi(t_E=0)=\phi}D\qty[\phi] e^{-I_E\qty[\phi]}\label{227intc}.
\end{align}

If we switch to polar coordinates, $\{t_E, x\} \rightarrow \{\theta, \rho\}$, the Hamiltonian, $H \rightarrow H_\theta$, will now be the generator of rotations, as we observed in \eqref{dthetadrot}. Therefore, instead of generating the translation of states from $t_E\rightarrow -\infty$ to $t_E=0$, it will now generate the rotation of states by a certain angle $\Delta\theta$. Also, we should note that the amplitude \eqref{227intc} occurs between states at $t=0$. Therefore, we have that $\Delta\theta=\pi$, and the rotation (counterclockwise) of states will occur from $- \pi$ to $0$ (see figure \ref{fig6c}). Hence,
\begin{align}
\phi(\theta=-\pi)&=\phi_L,\label{226=-pi}\\
\phi(\theta=0)&=\phi_R,\label{226=0}
\end{align}
where $\phi_R$ and $\phi_L$ correspond to the Rindler fields in regions I and IV of Minkowski, at $t=0$, respectively.\\

In addition, in order to perform the amplitude we must apply the CPT antiunitary operator, represented by $\Theta$ (see appendix \ref{Apendice5}) on the Rindler state in the left wedge, defined for $x<0$ and evolving in $t$ in reverse,
\begin{equation}
\Theta\phi_L\Theta^{-1}=\Tilde{\phi}_R\quad,\quad\Theta\ket{\phi_L}=\ket*{\Tilde{\phi}_R},
\end{equation}
in such a way that it evolves in space and time in the same way as $\ket{\phi_R}$,
\begin{equation}
    \bra{\phi}\ket{0}\propto\bra{\phi_R}e^{-\pi H_\theta}\ket*{\tilde{\phi}_R}=\bra{\phi_R}e^{-\pi H_\theta}\Theta\ket{\phi_L}\label{227amridi}.
\end{equation}

\subsubsection{Thermofield double state}

Inserting the identity operator in terms of the eigenstates of the Hamiltonian for $\phi_R$ in \eqref{227amridi}, knowing that $H_R = gH_\theta$, we have
\begin{align}
\bra{\phi_M}\ket{0_M}&\propto\bra{\phi_R}e^{-\frac{\pi}{g}H_R}\Theta\ket{\phi_L}\\
&\propto\displaystyle\sum_n\bra{\phi_R}e^{-\frac{\pi}{g}H_R}\ket{n_R}\bra*{n_R}\Theta\ket{\phi_L}.
\end{align}

Since $\Theta$ is an antiunitary operator, we have $\bra*{n_R}\Theta\ket{\phi_L}=\bra*{\phi_L}\Theta^{-1}\ket{n_R}$ (see equation \eqref{eqn190}). Then,
\begin{align}
\bra{\phi_M}\ket{0_M}&\propto\displaystyle\sum_n e^{-\frac{\pi}{g} E_n}\bra{\phi_R}\ket{n_R}\bra{\phi_L}\Theta^{-1}\ket*{n_R}.\label{phim0m}
\end{align}

The Hilbert spaces spanned by $\ket*{\Tilde{n}_L}=\Theta^{-1}\ket{n_R}$ and $\ket{n_R}$ are $\mathcal{H}_L$ and  $\mathcal{H}_R$, respectively. So, from \eqref{phim0m}, we notice that for $\ket{\phi_L}$ and $\ket{\phi_R}$, given in basis of $\ket*{\Tilde{n}_L}$ and $\ket{n_R}$, we have 
\begin{align}
\ket{\phi_M}\propto\ket{\phi_L}\otimes\ket{\phi_R},\\
\ket{0_M}\propto \displaystyle\sum_ne^{-\frac{\pi}{g} E_n}
\ket{n_R}\otimes \Theta^{-1}\ket*{n_R},
\end{align}
such that $\ket{0_M}$ and $\ket{\phi_M}$ are defined on $\mathcal{H}=\mathcal{H}_L \otimes \mathcal{H}_R$.\\

Applying the normalization condition $\bra{0}\ket{0}=1$, we obtain
\begin{align}
    1&=\abs{A}^2\sum_n\sum_m e^{-\frac{\pi}{g}\qty(E_n+E_m)}\qty[\bra{n_R}\otimes\qty(\Theta^{-1}\ket{n_R})^\dagger]\qty(\ket{m_R}\otimes\Theta^{-1}\ket{m_R})\\
    &=\abs{A}^2\sum_n\sum_m e^{-\frac{\pi}{g}\qty(E_n+E_m)}\braket{n_R}{m_R}\qty(\Theta^{-1}\ket{n_R})^\dagger\Theta^{-1}\ket{m_R}\\
    &=\abs{A}^2\sum_n\sum_m e^{-\frac{\pi}{g}\qty(E_n+E_m)}\delta_{n,m}\braket{m_R}{n_R}\\
    &=\abs{A}^2\sum_ne^{-\frac{2\pi}{g} E_n},
\end{align}
where we have used \eqref{sputi12}. Additionally, from \eqref{patichon}, the proportionality constant can be expressed as
\begin{equation}
    A=\qty(\sum_ne^{-\frac{2\pi}{g} E_n})^{-1/2}=\frac{1}{\sqrt{Z\qty(\frac{2\pi}{g})}},
\end{equation}
where
\begin{equation}
    Z(\beta)=\sum_n e^{-\beta E_n}.
\end{equation}

Therefore, the ground state for the scalar field in Minkowski space at $t=0$ is given by
\begin{equation}\label{Tfd2??}
\ket{0_M(\beta)} =\frac{1}{\sqrt{Z(\beta)}}\sum_{n}e^{-\frac{\beta}{2}E_n}\ket{n_R}\otimes\Theta^{-1}\ket{n_R},
\end{equation}
where $\beta=\frac{2\pi}{g}$.\\

The ground state \eqref{Tfd2??} is referred to as the thermofield double state, which is a temperature-dependent ``vacuum" state, as proposed by Umezawa and Takahashi \cite{Takahasi:1974zn,Takahashi:1996zn,Umezawa:1982nv,Matsumoto:1982ry} (see also \cite{maldacena2003eternal,israel1976thermo,harlow2016jerusalem}).\\

Before proceeding with our analysis of the density matrix, it is important to mention that, in addition to the thermofield double state $(TFD)$ formalism used in this review, other relevant approaches for analyzing thermal field theory include the Matsubara formalism \cite{matsubara1955new}, known as the ``imaginary time" formalism, in which time takes imaginary values ranging from $0$ to $-i\beta$, where $\beta$ is the inverse temperature. Additionally, there is the closed-time path $(CTP)$ formalism, also referred to as the ``real time" formalism, based on the Schwinger-Keldysh formulation \cite{schwinger1961brownian,keldysh2024diagram}, where, unlike $TFD$, both fields are physical. Finally, we also have the field theory in the Rindler frame formalism \cite{barman2024field}, which, at the non-interacting level and for non-analytic continuation, can fully mimic the $TFD$ formalism, with additional advantages such as both fields being part of the original system, while also encoding features of the $CTP$ formalism.

\subsubsection{Density matrix of the physical system}

The density matrix (appendix \ref{Apendice4}) for the pure ground state of the field theory in Minkowski, which is the thermofield double state, is given by
\begin{equation}
    \rho_M = \ket{0_M(\beta)}\bra{0_M(\beta)}\label{mdm}
\end{equation}

Thus,
\begin{align}
    \rho_M&=\displaystyle\sum_m\sum_{m'}\frac{e^{-\frac{\beta}{2}\qty(E_m+E_{m'})}}{Z(\beta)}\ket{m_R}\otimes\Theta^{-1}\ket{m_R}\bra*{m'_R}\otimes\qty(\Theta^{-1}\ket{m'_R})^\dagger\nonumber\\
&=\displaystyle\sum_m\sum_{m'}\frac{e^{-\frac{\beta}{2}\qty(E_m+E_{m'})}}{Z(\beta)}\ket{m_R}\otimes\ket*{\Tilde{m}_L}\bra*{m'_R}\otimes\bra*{\Tilde{m}'_L}\label{rhomcomp},
\end{align}
where $\ket*{\Tilde{m}_L}=\Theta^{-1}\ket{m_R}$.\\

The reduced density matrix for the Rindler field in the right wedge, $\phi_R$, which describes the physical system for an observer undergoing constant acceleration in Minkowski spacetime, is obtained by taking the partial trace of the contribution from $\phi_L$ in \eqref{rhomcomp},
\begin{equation}\label{mdrr}
\rho_{R}=\Tr_{L}\rho_M\\
=\sum_n \bra*{\Tilde{n}_L} \rho_M \ket*{\Tilde{n}_L}.
\end{equation}

From \eqref{rhomcomp} in \eqref{mdrr},
\begin{equation}
\rho_R=\sum_n\sum_m\sum_{m'}\frac{e^{-\frac{\beta}{2}\qty(E_m+E_{m'})}}{Z(\beta)}\ket{m_R}\bra{m'_R}\delta_{nm'}\delta_{mn}.
\end{equation}

Hence,
\begin{align}
    \rho_R&=\displaystyle\sum_n \frac{e^{-\beta E_n}}{Z(\beta)}\ket{n_R}\bra{n_R}\\
&=\frac{e^{-\beta H_R}}{Z(\beta)}\label{rhoR269}.
\end{align}

We note that the density matrix $\rho_R$ describes a mixed ensemble in thermal equilibrium with a heat bath at a temperature given by $T=\beta^{-1}=g/2\pi$, which is consistent with what was obtained in \eqref{215temunru}. Consequently, the thermofield double state \eqref{Tfd2??} is an entangled state of the Hamiltonian eigenstates $\qty{\ket{n_R},\ket*{\Tilde{n}_L}=\Theta^{-1}\ket{n_R}}$.\\

Let $N_R$ be the number operator for a single mode with energy $\Omega$ (recall that the scalar field is massless, so the relationship between energy and momentum is given by $\Omega = \abs{k}$), with eigenstate $\ket{n_R}$, such that  
\begin{equation}  
N_R\ket{n_R} = n \Omega \ket{n_R}.  
\end{equation}  
Then, the mean number of particles at energy $\Omega$ is given by
\begin{align}
    n_R(\Omega)&=\Tr N_{R}\rho_R=\frac{1}{Z(\beta)}\sum_{n} \bra{n_R}N_R\,e^{-\beta H_{R}}\ket{n_R}\\
&=\frac{\displaystyle\sum_{n}n e^{-\beta n\Omega}}{\displaystyle\sum_{n} e^{-\beta n\Omega}},
\end{align}
where $H_R\ket{n_R}=n\Omega\ket{n_R}$. Thus, since $n$ is the number of quanta at energy $\Omega$, we perform the sum from $0$ to $\infty$:
\begin{equation}
n_R(\Omega)=\frac{1}{e^{\beta \Omega}-1},
\end{equation}
where $\beta=1/T=2\pi/g$. This result is consistent with \eqref{db-ei}.

\section{Conclusions}
In this review, it was shown that the extended geometry of Rindler spacetime, contained within region I of Minkowski spacetime (see figure \ref{fig1}) and referred to as the right Rindler wedge, corresponds precisely to Minkowski spacetime, which includes a time-reversed copy in region IV, known as the left Rindler wedge. Notably, this geometry exhibits a structure analogous to that of the eternal black hole, as confirmed through an analysis of the causal structure. Furthermore, since the metrics in Minkowski and Rindler spacetimes are related through a conformal transformation, it was possible to express the Fourier modes of one reference frame in terms of the other using the Bogoliubov-Valatin transformation. As a result, from the perspective of the accelerated observer, the Minkowski vacuum state is described as a mixed ensemble characterized by Bose-Einstein statistics with a temperature proportional to the acceleration of the non-inertial observer. Additionally, an analogy with Schwarzschild black hole was established, showing how the Unruh effect relates to the thermal radiation observed around such black hole. We also analyzed the physics of a particle detector (under constant acceleration) coupled to the scalar field, which provides another manifestation of the Unruh effect. Finally, it was demonstrated that the ground state of the theory corresponds to the thermofield double state, which represents an entangled system of energy eigenstates in the left and right wedges.\\

\acknowledgments
The author would like to thank Professor Carlos R. Ordóñez, for insightful discussions on the elaboration of figures \ref{fig5a}, \ref{fig}, \ref{fig6a}, and \ref{fig15}, which were originally conceived for use in \cite{chakraborty2024path}, as well as on the section regarding the Unruh-DeWitt detector. The author also wishes to express special thanks to the Center for Mexican American and Latino/a Studies at the University of Houston for their generous support through a Lydia Mendoza Fellowship. He was also partially supported by the Army Research Office (ARO), Grant W911NF-23-1-0202.

\appendix
\section{Null coordinates}\label{Apendice1}

For a spacetime $\qty{t,x}$ with a flat metric (or conformally flat), the null surfaces are
\begin{equation}
x\pm t= \alpha,
\end{equation}
where $\alpha$ is a constant. Let us define the following surface functions:
\begin{align}
v(t,x)=t+x+\alpha,\\
u(t,x)=t-x+\alpha.
\end{align}

The normal vectors to these surfaces are obtained by applying the gradient:
\begin{itemize}
    \item[a.] For $v$:
\begin{align}
v_\mu=\partial_\mu v&\equiv \left(\frac{\partial}{\partial t},\frac{\partial}{\partial x} \right)v \nonumber\\
&\equiv (1,1).
\end{align}
\item[b.] For $u$:
\begin{align}
u_\mu=\partial_\mu u &\equiv \left(\frac{\partial}{\partial t},\frac{\partial}{\partial x} \right)u \nonumber\\
&\equiv (1,-1).
\end{align}
\end{itemize}

As we can see, normal vectors to null surfaces are null vectors:
\begin{align}
v^\mu v_\mu=0,\\
u^\mu u_\mu=0.
\end{align}

So, $u$ and $v$ are called null coordinates.

\section{Generators of the Lorentz Group}\label{Apendice3}
The infinitesimal coordinate variation obtained from the Lorentz transformation,
\begin{equation}
    {x'}^{\mu}= {\Lambda^\mu}_\nu x^{\nu} \approx \qty({\delta^\mu}_\nu + {\omega^\mu}_\nu)x^\nu\label{infi123},
\end{equation}
is given by
\begin{equation}
    \delta x^\mu = x'^\mu-x^\mu\approx{\omega^\mu}_\nu x^\nu\label{301}.
\end{equation}

Additionally, by utilizing the invariance of the spacetime interval,
\begin{align}
    \eta_{\mu\nu}x^{\mu}x^\nu&=\eta_{\alpha\beta}x'^{\alpha}x'^\beta\nonumber\\
    &=  \eta_{\alpha\beta}{\Lambda^\alpha}_{\mu} x^\mu{\Lambda^\beta}_{\nu} x^\nu,
\end{align}
we obtain
\begin{equation}
    \eta_{\mu\nu}= \eta_{\alpha\beta}{\Lambda^{\alpha}}_\mu{\Lambda^\beta}_{\nu}.
\end{equation}

Applying the infinitesimal transformation \eqref{infi123},
\begin{align}
    \eta_{\mu\nu}&\approx\eta_{\alpha\beta}\qty({\delta^\alpha}_\mu + {\omega^\alpha}_\mu)({\delta^\beta}_\nu + {\omega^\beta}_\nu)\nonumber\\
    &\approx\eta_{\mu\nu}+\omega_{\nu\mu}+\omega_{\mu\nu},
\end{align}
we obtain that $\omega_{\mu\nu}$ is antisymmetric
\begin{equation}
    \omega_{\mu\nu}=-\omega_{\nu\mu}\label{antisy123}.
\end{equation}

Doing some manipulations on \eqref{301},
\begin{align}
    \delta x^\mu &\approx {\omega^\mu}_\nu x^\nu\nonumber\\
    &\approx \omega^{\mu\nu} x_\nu=\omega^{\alpha\nu} x_\nu {\delta^\mu}_\alpha\nonumber\\
    &\approx\omega^{\alpha\nu} x_\nu \partial_\alpha x^\mu,
\end{align}
and using \eqref{antisy123},
\begin{align}
    \delta x^\mu &\approx \omega^{\alpha\nu} \frac{1}{2}\qty(x_\nu \partial_\alpha - x_\alpha \partial_\nu) x^\mu\nonumber\\
    &\approx -\frac{1}{2} \omega_{\alpha\nu}M^{\alpha\nu}x^\mu,
\end{align}
we obtain the generator of transformations for the Lorentz group
\begin{equation}
    M^{\alpha\nu}=x^\alpha\partial^\nu-x^\nu\partial^\alpha.
\end{equation}

Finally, the infinitesimal generator of boosts in the $i$-direction is given by
\begin{equation}
    M^{0i} = x^0 \partial^i - x^i \partial^0 = t \partial_i + x^i \partial_t,
\end{equation}
while the infinitesimal generator of spatial rotations is given by
\begin{equation}
    M^{ij} = x^i \partial^j - x^j \partial^i = \epsilon^{ij}_k J^k.
\end{equation}

\section{Properties of the conformal transformation of the metric}\label{Apendice2}

Let us consider the line element
\begin{equation}\label{1.1.3 ds21}
ds^2=g_{\mu\nu}dx^{\mu}dx^{\nu},
\end{equation}
and define the one conformal to \eqref{1.1.3 ds21} given by the metric $\tilde{g}_{\mu\nu}$,
\begin{align}
\tilde{g}_{\mu\nu} &= \Omega^2g_{\mu\nu}\label{1.1.3 gtil},\\
d\tilde{s}^2 &= \tilde{g}_{\mu\nu}dx^\mu dx^\nu = \Omega^2 g_{\mu\nu}dx^\mu dx^\nu \nonumber\\
&= \Omega^2 ds^2,
\end{align}
such that, the conformal factor $\Omega$ is a positive-definite and smooth function of coordinates.\\

In addition, we have the following properties for the conformal transformation:
\begin{itemize}
    \item[a.] Does not preserve vector's length, but it does maintain its type:
    \begin{equation}
\tilde{g}_{\mu\nu}A^\mu A^\nu=\Omega^2 g_{\mu\nu}A^\mu A^\nu=\Omega^2A^\mu A_\mu.
\end{equation}

From this result, it is evident that the length is not preserved. Furthermore, as $\Omega^2$ is positive-definite, it will not change the vector's type:
\begin{itemize}
\item Timelike: $A^\mu A_\mu < 0 \rightarrow \Omega^2A^\mu A_\mu <0$.
\item Spacelike: $A^\mu A_\mu > 0 \rightarrow \Omega^2A^\mu A_\mu >0$.
\item Lightlike: $A^\mu A_\mu = 0 \rightarrow \Omega^2A^\mu A_\mu =0$.
\end{itemize}

\item[b.] Preserves the angle between vectors:

In Euclidean signature,
\begin{align}
x^\mu= (\Vert x  \Vert \sin{\beta} , \Vert x \Vert \cos{\beta} ),
\end{align}
where $\Vert x \Vert =\sqrt{x^\mu x_\mu}$.\\

Let us take the vectors $A^\mu$ and $B^\mu$, such that $\theta$ is the angle between them:
\begin{align}
A^\mu=(\Vert A \Vert \sin{(\alpha+\theta)} , \Vert A \Vert\cos{(\alpha+\theta)}),\\
B^{\mu}=(\Vert B \Vert \sin{(\alpha)} , \Vert B \Vert\cos{(\alpha)} ).
\end{align}

From their product, we obtain
\begin{align}
\cos{(\theta)} = \frac{g_{\mu\nu}A^\mu B^{\nu}}{\Vert A \Vert \Vert B \Vert}.
\end{align}

Then, the conformal transformation $\tilde{g}_{\mu\nu}=\Omega(x)^2g_{\mu\nu}$ gives us
\begin{align}
\cos{(\tilde{\theta})} = \frac{\tilde{g}_{\mu\nu}A^\mu B^{\nu}}{\Vert \tilde{A} \Vert \Vert \tilde{B} \Vert}=\frac{\Omega(x)^2g_{\mu\nu}A^\mu B^{\nu}}{\Omega(x) \Vert A \Vert \Omega(x) \Vert B \Vert}.
\end{align}

Therefore,
\begin{equation}
\cos{(\tilde{\theta})}=\cos{(\theta)}.
\end{equation}

\item[c.] Preserves null geodesics:

We have the Christoffel symbol for the metric $\tilde{g}_{\mu\nu}$:
\begin{equation}\label{cris}
\tilde{\Gamma}^\alpha_{\mu\nu}=\frac{1}{2}\tilde{g}^{\alpha\beta}\qty(\tilde{g}_{\beta\mu,\nu}+\tilde{g}_{\beta\nu,\mu}-\tilde{g}_{\mu\nu,\beta}).
\end{equation}

The geodesic equation is given by
\begin{equation}\label{geode1}
\frac{d^2x^\alpha}{d\tilde{\omega}^2}+\tilde{\Gamma}^\alpha_{\mu\nu}\frac{dx^\mu}{d\tilde{\omega}}\frac{dx^\nu}{d\tilde{\omega}}=0.
\end{equation}

Let us define the metric $\tilde{g}^{\mu\nu}$ as conformal to $g^{\mu\nu}$, i.e., $\tilde{g}_{\mu\nu}=\Omega(x)^{2}g_{\mu\nu}$. Then,
\begin{equation}\label{cris2}
\tilde{\Gamma}^\alpha_{\mu\nu}=\Gamma^\alpha_{\mu\nu}+\delta^\alpha_\mu\Omega^{-1}\partial_\nu\Omega+\delta^\alpha_\nu\Omega^{-1}\partial_\mu\Omega-g_{\mu\nu}\Omega^{-1}\partial^{\alpha}\Omega.
\end{equation}

Using the above result in \eqref{geode1},
\begin{multline}
\frac{d^2x^\alpha}{d\tilde{\omega}^2}+\tilde{\Gamma}^\alpha_{\mu\nu}\frac{dx^\mu}{d\tilde{\omega}}\frac{dx^\nu}{d\tilde{\omega}}=\frac{d^2x^\alpha}{d\tilde{\omega}^2}+\Gamma^\alpha_{\mu\nu}\frac{dx^\mu}{d\tilde{\omega}}\frac{dx^\nu}{d\tilde{\omega}}+\\+2\Omega^{-1}\partial_\nu\Omega\frac{dx^\alpha}{d\tilde{\omega}}\frac{dx^\nu}{d\tilde{\omega}}-g_{\mu\nu}\frac{dx^\mu}{d\tilde{\omega}}\frac{dx^\nu}{d\tilde{\omega}}\Omega^{-1}\partial^{\alpha}\Omega=0.
\end{multline}

Taking the affine parametrization between $\tilde{\omega}$ and $\omega$, in $\tilde{g}_{\mu\nu}$ and $g_{\mu\nu}$, respectively,
\begin{equation}
d\tilde{\omega}=\Omega^2d\omega,
\end{equation}
we obtain
\begin{multline}\label{b18}
\frac{d^2x^\alpha}{d\tilde{\omega}^2}+\tilde{\Gamma}^\alpha_{\mu\nu}\frac{dx^\mu}{d\tilde{\omega}}\frac{dx^\nu}{d\tilde{\omega}}=\Omega^{-4}\left[ \frac{d^2x^\alpha}{d\omega^2}+\Gamma^\alpha_{\mu\nu}\frac{dx^\mu}{d\omega}\frac{dx^\nu}{d\omega}+2\Omega^{-1}\partial_\nu\Omega\frac{dx^\alpha}{d\omega}\frac{dx^\nu}{d\omega}+\right.\\\left.+\,\Omega^2\frac{d\Omega^{-2}}{d\omega}\frac{dx^\alpha}{d\omega}-g_{\mu\nu}\frac{dx^\mu}{d\omega}\frac{dx^\nu}{d\omega}\Omega^{-1}\partial^{\alpha}\Omega\right] =0.
\end{multline}

From the chain rule we have
\begin{equation}\label{Ap299}
2\Omega^{-1}\partial_\nu\Omega\frac{dx^\alpha}{d\omega}\frac{dx^\nu}{d\omega}+\Omega^2\frac{d\Omega^{-2}}{d\omega}\frac{dx^\alpha}{d\omega}=0.
\end{equation}

Applying \eqref{Ap299} in \eqref{b18}, we find
\begin{multline}
\frac{d^2x^\alpha}{d\tilde{\omega}^2}+\tilde{\Gamma}^\alpha_{\mu\nu}\frac{dx^\mu}{d\tilde{\omega}}\frac{dx^\nu}{d\tilde{\omega}}=\Omega^{-4}\left[ \frac{d^2x^\alpha}{d\omega^2}+\Gamma^\alpha_{\mu\nu}\frac{dx^\mu}{d\omega}\frac{dx^\nu}{d\omega}-g_{\mu\nu}\frac{dx^\mu}{d\omega}\frac{dx^\nu}{d\omega}\Omega^{-1}\partial^{\alpha}\Omega\right] =0.
\end{multline}

As we know, for null geodesics
\begin{equation}
g_{\mu\nu}\frac{dx^\mu}{d\omega}\frac{dx^\nu}{d\omega}=0.
\end{equation}

Then,
\begin{equation}
    \frac{d^2x^\alpha}{d\tilde{\omega}^2}+\tilde{\Gamma}^\alpha_{\mu\nu}\frac{dx^\mu}{d\tilde{\omega}}\frac{dx^\nu}{d\tilde{\omega}}=\Omega^{-4}\left[ \frac{d^2x^\alpha}{d\omega^2}+\Gamma^\alpha_{\mu\nu}\frac{dx^\mu}{d\omega}\frac{dx^\nu}{d\omega}\right] =0.
\end{equation}

As $\Omega$ is defined positive and smooth, if $x^\mu$, with parameter $\tilde{\omega}$, is a null geodesic in $\tilde{g}_{\mu\nu}$, it is also null in $g_{\mu\nu}$, with $\omega$ as parameter. Therefore, the conformal transformation preserves the causal structure.
\end{itemize}

\section{Pure, mixed and entangled states}\label{Apendice4}
\addtocontents{toc}{\setcounter{tocdepth}{-1}}
According to Sakurai and Napolitano \cite{sakurai2014modern}.
\subsection{Pure state}
A pure ensemble is a collection of physical systems such that every member is characterized by the same ket. In other words, the ensemble is described by a single vector $\ket{\psi}$ in Hilbert space. The density matrix for a pure state is given by
\begin{equation}
\rho=\ket{\psi}\bra{\psi},
\end{equation}
and satisfies the following properties:
\begin{align}
\rho^2&=\rho,\\
\tr\rho^2&=1.
\end{align}

\subsection{Mixed state}
A mixed ensemble can be viewed as a probabilistic mixture of pure states $\ket{\psi_i}$. In addition, it cannot be described by a single vector, only by a density matrix,
\begin{equation}
\rho=\displaystyle\sum_i p_i \ket{\psi_i}\bra{\psi_i},
\end{equation}
where
\begin{equation}
    \displaystyle\sum_i p_i=1.
\end{equation}

The density matrix for a mixed state satisfies the following properties:
\begin{align}
\rho^2&\neq\rho,\\
\tr\rho^2&<1.
\end{align}

\subsection{Reduced density matrix}
Considering the state $\ket{\psi_{AB}}$ defined in the Hilbert space,
\begin{equation}
\mathcal{H}=\mathcal{H}_A\otimes\mathcal{H}_B,
\end{equation}
we can obtain the density matrix for each subsystem, known as the reduced density matrix, by taking the partial trace over the subsystem we want to exclude from $\rho_{AB}$. For example, the density matrix defined on $\mathcal{H_A}$ is given by
\begin{equation}
\rho_A=\Tr_B\qty[ \rho_{AB}].
\end{equation}

\subsubsection{Product States}\label{apEespro}  
If the pure state $\ket{\psi_{AB}}$ can be expressed as the tensor product of the system states defined in $\mathcal{H}_A$ and $\mathcal{H}_B$,
\begin{equation}\label{311rr}
\ket{\psi_{AB}}=\ket{\psi_A}\otimes\ket{\psi_B},
\end{equation}
then the reduced density matrix for the subsystem $A$, given by
\begin{align}
\rho_A &= \bra{\psi_B}\rho_{AB}\ket{\psi_B} = \ket{\psi_{A}}\bra{\psi_{A}},
\end{align}
is a pure ensemble (the same holds for subsystem $B$).\\

\subsubsection{Entangled States}\label{apEesent}  
If the pure state $\ket{\psi_{AB}}$ cannot be expressed as a product of states, i.e.,
\begin{equation}
\ket{\psi_{AB}} \neq \ket{\psi_A}\otimes\ket{\psi_B},
\end{equation}
then it is an entangled state.\\

For example,
\begin{align}
    \ket{\psi_{AB}} &= \frac{1}{\sqrt{2}} \qty[\ket{0_A}\otimes\ket{1_B}+\ket{1_A}\otimes\ket{0_B}]\label{319rr},
\end{align}
is an entangled state of the basis kets of subsystems $A$ and $B$, as it cannot be written as a product of the states $\ket{\psi_A}$ and $\ket{\psi_B}$,
\begin{align}
    \ket{\psi_A} = \alpha_1\ket{0_A} + \beta_1\ket{1_A},\\
    \ket{\psi_B} = \alpha_2\ket{0_B} + \beta_2\ket{1_B},
\end{align}
for any possible values of $\alpha_{1,2}$ and $\beta_{1,2}$.\\ 

Therefore, the reduced density matrices for subsystems $A$ and $B$ will describe mixed ensembles. From the previous example, we have:
\begin{align}
    \rho_A &= \frac{1}{2} \qty[ \ket{0_A}\bra{0_A} + \ket{1_A}\bra{1_A}]\nonumber\\
    &= \displaystyle\sum_{i=0}^1 p_i\ket{i_A}\bra{i_A},
\end{align}
where $p_0 = p_1 = \frac{1}{2}$. The result is identical for subsystem $B$.\\
 
Finally, we have:
\begin{align}
    \ket{\psi_{AB}}\,\,\text{is entangled} &\Leftrightarrow \rho_A\,\,\text{is mixed},\\
    \ket{\psi_{AB}}\,\,\text{is not entangled} &\Leftrightarrow \rho_A\,\,\text{is pure}.
\end{align}

\addtocontents{toc}{\setcounter{tocdepth}{2}}
\section{Antiunitary operator}\label{Apendice5}
\addtocontents{toc}{\setcounter{tocdepth}{-1}}
According to Sakurai and Napolitano \cite{sakurai2014modern}.\\

Let the transformations be given by the operator $\Theta$ on the states $\ket{\psi}$ and $\ket{\phi}$,
\begin{equation}
    \ket*{\Tilde{\psi}}=\Theta\ket{\psi},\quad \ket*{\Tilde{\phi}}=\Theta\ket{\phi}.
\end{equation}

We say that the operator $\Theta$ is antiunitary if it satisfies:
\begin{align}
    \bra*{\Tilde{\phi}}\ket*{\Tilde{\psi}}&=\bra{\psi}\ket{\phi}\label{e2222},\\
    \Theta\qty(c_1\ket{\psi}+c_2\ket{\phi})&=c_1^*\ket*{\Tilde{\psi}}+c_2^*\ket*{\Tilde{\phi}}.\label{e3333}
\end{align}

The antiunitary operator $\Theta$ can be written as the product of a unitary operator $U$ and a complex-conjugate operator $K$,
\begin{equation}
    \Theta=UK.
\end{equation}

The operator $K$ transforms any coefficient that multiplies a ket (it acts to the right) into its complex conjugate, while leaving the basis kets unaffected:
\begin{align}
    K\ket{\psi}&=K\sum_a \psi_a\ket{a}\nonumber\\
    &=\sum_a \psi_a^*K\ket{a}\nonumber\\
    &=\sum_a \psi_a^*\ket{a},
\end{align}
we note that since the basis ket can be expressed as a column vector with real entries given by zeros and ones, the application of the operator $K$ has no effect.\\

In this way, we have
\begin{align}
    \Theta\qty(c_1\ket{\psi}+c_2\ket{\phi})&=UK\qty(c_1\ket{\psi}+c_2\ket{\phi})=c_1^*UK\ket{\psi}+c_2^*UK\ket{\phi}\nonumber\\
    &=c_1^*\Theta\ket{\psi}+c_2^*\Theta\ket{\phi}\nonumber\\
    &=c_1^*\ket*{\Tilde{\psi}}+c_2^*\ket*{\Tilde{\phi}},
\end{align}
which satisfies condition \eqref{e3333}.\\

The definition of the adjoint involves an extra complex conjugation
\begin{align}
    \Theta\ket{\psi}&=\Theta\sum_a \psi_a\ket{a}\nonumber\\
    &=\sum_a\psi_a^*\Theta\ket{a}\nonumber\\
    &=\sum_a\psi_a^* U\ket{a}.
\end{align}
Thus,
\begin{align}
    (\Theta\ket{\psi})^\dagger=\sum_a\psi_a \bra{a}U^\dagger.
\end{align}

Let's calculate $\bra*{\Tilde{\psi}}\ket*{\Tilde{\phi}}$, knowing that $\ket{a}$ and $\ket{b}$ are basis kets such that $\bra{a}\ket{b}=\delta_{a,b}$:
\begin{align}
\bra*{\Tilde{\psi}}\ket*{\Tilde{\phi}}&=(\Theta\ket{\psi})^\dagger\Theta\ket{\phi}\nonumber\\
&=\sum_{a,b}\psi_a \bra{a}U^\dagger \phi_b^* U\ket{b}=\sum_{a,b}\psi_a\phi_b^*\delta_{a,b}\nonumber\\
    &=\bra{\phi}\ket{\psi},
\end{align}
which satisfies condition \eqref{e2222}.\\

Let's analyze the action of the operator $\Theta^{-1}$, such that $\ket{\psi'}=\Theta^{-1}\ket{\psi}$. Starting from $\ket{\psi}=\Theta^{-1}\Theta\ket{\psi}$ we have
\begin{align}
    \ket{\psi}=\Theta^{-1}\Theta\ket{\psi}=\Theta^{-1}\sum_a\psi_a^* U\ket{a}=\sum_a K^{-1}U^\dagger\psi_a^* U\ket{a}=\sum_a K^{-1}\psi_a^* \ket{a}.
\end{align}

Knowing that $\ket{\psi}=\sum_a \psi_a\ket{a}$, we have that $K^{-1}\psi_a^*\ket{a}=\psi_a\ket{a}$. In this way, we note that the action of $K^{-1}$ on the coefficients of the ket is to transform them into their complex conjugate, and like $K$, it leaves the basis ket $\ket{a}$ unchanged. Therefore,
\begin{equation}
    \Theta^{-1}\psi_a\ket{a}=\psi_a^*\Theta^{-1}\ket{a},
\end{equation}
with which $\Theta^{-1}$ satisfies condition \eqref{e3333}.\\

Likewise, we have that $\Theta^{-1}\ket{a}=K^{-1}U^\dagger\ket{a}$. Considering that the action of $U^\dagger\ket{a}$ produces a new set of orthonormal bases $\ket{a_u}=U^\dagger\ket{a}$, such that $\bra{a_u}\ket{b_u}=\delta_{a,b}$. Therefore, $K^{-1}U^\dagger\ket{a}=U^\dagger\ket{a}$. In this way, we have
\begin{align}
\Theta^{-1}\ket{\psi}=\sum_a\psi_a^*U^\dagger\ket{a}\Rightarrow \qty(\Theta^{-1}\ket{\psi})^\dagger=\sum_a\psi_a\bra{a}U.
\end{align}

From the previous result, let's calculate $\bra{\psi'}\ket{\phi'}$ knowing that $\ket{a}$ and $\ket{b}$ are basis kets such that $\bra{a}\ket{b}=\delta_{a,b}$:
\begin{align}
\bra{\psi'}\ket{\phi'}&=\qty(\Theta^{-1}\ket{\psi})^\dagger \Theta^{-1}\ket{\phi}\nonumber\\
    &=\sum_{a,b}\psi_a\bra{a}U \phi_b^*U^\dagger\ket{b}\nonumber\\
    &=\sum_{a,b}\psi_a\phi_b^*\delta_{a,b}\nonumber\\
    &=\bra{\phi}\ket{\psi},\label{sputi12}
\end{align}
with which $\Theta^{-1}$ satisfies condition \eqref{e2222}.\\

Thus, if $\Theta$ is an antiunitary operator, $\Theta^{-1}$ is also antiunitary.\\

Finally, from $\ket*{\Tilde{\psi}}=\Theta\ket{\psi}$, we have $\ket{\psi}=\Theta^{-1}\ket*{\Tilde{\psi}}$. Thus:
\begin{equation}
    \bra*{\Tilde{\psi}}\Theta\ket{\phi}=\bra{\phi}\Theta^{-1} \ket*{\Tilde{\psi}}.\label{eqn190}
\end{equation}

\newpage

\bibliographystyle{JHEP}
\bibliography{biblio}

\end{document}